\documentclass[resetfootnote,twocolumn]{aastex701}
\usepackage{siunitx}
\newcommand{\ppxf}{\texttt{pPXF}}
\newcommand{\marz}{\texttt{Marz}}

\usepackage{graphicx}
\usepackage{amsmath}
\usepackage{mathtools}
\usepackage{amssymb}
\newcommand{\code}[1]{\texttt{#1}}
\newcommand{\urgent}[1]{{\textcolor{black}{#1}}}
\newcommand{\katya}[1]{{\textcolor{black}{#1}}}
\newcommand{\eric}[1]{{\textcolor{black}{#1}}}

\newcommand{\moreurgent}[1]{{\textcolor{black}{#1}}}
\usepackage{booktabs}

\shorttitle{NGC 253 Stellar Halo Kinematics}

\begin{document}

\title{She Sees Shells by the Stellar Shore: A Kinematic View of NGC 253's Stellar Halo\footnote{This paper includes data gathered with the 6.5 meter Magellan Telescopes located at Las Campanas Observatory, Chile.}}

\author[0000-0003-2294-4187]{Katya Gozman}
\affiliation{Department of Astronomy, University of Michigan, 1085 S. University Ave, Ann Arbor, MI 48109-1107, USA}
\email{kgozman@stanford.edu}

\author[0000-0002-5564-9873]{Eric F.\ Bell}
\affiliation{Department of Astronomy, University of Michigan, 1085 S. University Ave, Ann Arbor, MI 48109-1107, USA}
\email{ericbell@umich.edu}

\author[0000-0001-9269-8167]{Richard D'Souza}
\affiliation{Vatican Observatory, Specola Vaticana, V-00120, Vatican City State}
\email{rdsouza@specola.va}

\author[0000-0002-3856-232X]{Mario Mateo}
\affiliation{Department of Astronomy, University of Michigan, 1085 S. University Ave, Ann Arbor, MI 48109-1107, USA}
\email{mmateo@umich.edu}

\author[0000-0002-4272-263X]{John I. Bailey III}
\affiliation{Caltech Optical Observatories, California Institute of Technology, Pasadena, CA 91125, USA}
\email{jibailey@caltech.edu}

\author[0000-0003-2325-9616]{Antonela Monachesi}
\affiliation{Departamento de Astronom\'{i}a, Universidad de La Serena, Av. Ra\'{u}l Bitr\'{a}n 1305, La Serena, Chile}
\email{amonachesi@userena.cl}

\author[0009-0006-2180-4309]{Camila Beltrand}
\affiliation{Departamento de Astronom\'{i}a, Universidad de La Serena, Av. Ra\'{u}l Bitr\'{a}n 1305, La Serena, Chile}
\affiliation{Facultad de Ingeniería y Arquitectura, Universidad Central de Chile, Av. Francisco de Aguirre 0405, La Serena, Chile.}
\email{camila.beltrand@userena.cl}





\begin{abstract}

The stellar halos of nearby galaxies are fossil records of ancient mergers, providing insight into the formation and growth of these systems. These halos often contain substructure in the form of streams, shells, and debris fields that make up dynamically young remnants of past accretions. These remnants can be probed with resolved-star spectroscopic observations, giving insight into the kinematics of bygone satellite companions. 
Such observations have been made for the Milky Way (MW) and Andromeda, but are much more challenging and expensive to do outside the Local Group. The Local Volume has an abundance of MW-mass galaxies with richly structured halos, such as NGC 253. Leveraging the power of Magellan's M2FS fiber-fed multiobject spectrograph, we have obtained medium-resolution spectra of 225 stars centered on NGC 253's southwestern shell. These spectra enabled us to make the first-ever measurement of the resolved-star stellar kinematics for 89 stars in NGC 253's halo. We also detect substructure in several ways and identify regions likely to be accreted and \textit{in-situ}, finding among them a low-velocity region coincident with the previously reported SW shell we believe was accreted recently, and an \textit{in-situ} component that traces the outer disk. After subtracting NGC 253's systemic velocity, we measure median line-of-sight velocities of $-34_{-12}^{+16}$ km/s and $167_{-14}^{+16}$ km/s for the accreted and \textit{in-situ} populations, respectively. This work highlights the utility of kinematics to differentiate between accreted and \textit{in-situ} stars, and acts as a pathfinder for future observations with next-generation 30 m telescopes and wide-field multiobject spectrographs. 

\end{abstract}

\keywords{\uat{Galaxies}{573} --- \uat{Stellar astronomy}{1583} --- \uat{Galaxy stellar halos}{598} }


\section{Introduction}\label{sec:intro}

One of the most important processes that govern the evolution of galaxies is merging. This so-called hierarchical model of galaxy formation is a direct prediction of the $\Lambda$ cold dark matter ($\Lambda$CDM) cosmological paradigm, wherein dark matter halos and their host galaxies grow over time through continuous accretion of smaller halos \citep{searlezinn1978ApJ_hierarchicalhaloformation, bullock2001, bullockjohnson, cooper2010}. Accreted stars and other tidal debris from merging satellite galaxies are deposited in the outskirts of their host galaxy, creating a faint and diffuse stellar halo that provides an archaeological record of the accretion histories of Milky Way (MW)-mass galaxies. The stellar halos of these galaxies contain signatures of different epochs of accretion: phase-mixed \textit{ex-situ} stars from old mergers, and substructures in the form of stellar streams, shells, and overdensities from more recent and large interactions, as well as satellite galaxies that may have been delivered by past merging partners \citep[e.g.,][]{merrifiendandkuijken1998, bullock2001, ibata2004MNRAS_M31GSSkinematics, somervile+dave_2015ARA&A}. 

Unfortunately, most of the substructures in a stellar halo have very low surface brightnesses ($\mu_V \geq 28$ mag/arcsec$^2$, \citealt{bullockjohnson, veracasanova2022MNRAS_stellarstreamsaccretionhistorieslink}) and large extent ($\sim 100-200$ kpc), making them difficult to detect, especially outside of the MW. Concerted efforts have been made to search for these overdensities in integrated light with deep imaging surveys \citep[e.g.][]{Martinez-Delgado2010, vandokkum2014ApJ_dragonflym101, merritt2016ApJ_dragonflyhalos, zaritsky2024AJ_lights, romanowskyr2025RNAAS_rubinm61stellarstream}, but this method is challenged by scattered light and regions of heavy Galactic cirrus. To circumvent this, resolving individual stellar populations in deep wide-field observations enables us to probe deeper surface brightnesses (down to $\sim 33$ mag/arcsec$^2$) and understand the specific stellar properties of the stars associated with different halo substructure.\footnote{Henceforth, ``halo" will refer specifically to the stellar halo of a galaxy unless otherwise specified.
}

Over the last decade, deep resolved-star photometric studies of stellar halos have been pioneered by surveys such as GHOSTS and individual imaging of select galaxies \citep[e.g.,][]{radburn-smith2011ApJS_ghosts_ngc253dist, monachesi2013ApJ_ghostsm81color, Monachesi2016, rejkuba2014, Harmsen2017,jang2020A&A_m101halo, smercina2020m81, gozman2023ApJ...947...21G}. \katya{These surveys have used halo populations to make important inferences about the dominant mergers of nearby galaxies, assuming that the majority of the halo mass comes from their most massive merger partners \citep{monachesi2019}. But stellar halos can contain stars that have been accreted through mergers, as well as \textit{in-situ} stars --- ones that formed within the main body of the galaxy but can be found at large radii due to different physical mechanisms \citep{bell2008, purcell2010MNRAS_heateddiskhalo, cooper2015MNRAS_insituhalo, monachesi2019, font2020MNRAS_ARTMESIsimshalos}. Disentangling accreted and \textit{in-situ} populations has been a major challenge for resolved stellar population studies of halos. Often, they can be partially distinguished by their spatial locations and stellar populations --- accreted stars are more likely to be found in the outer parts of the halo (dominating beyond $\sim 20-30$ kpc) and along their minor axes  and are generally more metal-poor with different chemical abundances than \textit{in-situ} stars, which are more likely to be metal-richer and predominantly found in the inner halo and near the disk plane \citep{zolotov2009ApJ_dualoriginhalos, cooper2015MNRAS_insituhalo, pillepich2015ApJ_spiralinsituexsitu, khoperskov2023A&A_HESTIA_insitu}. But the distinction is not always clear-cut, as simulations differ in their predictions of \textit{in-situ} and accreted mass fractions \citep{cooper2010,  font2011MNRAS_simshalosdiscs, tissera2014MNRAS_halos, monachesi2019} and they can be difficult to distinguish even in the nearest halos such as M31 \citep{escala2020ApJ_M31innerhalo} without multidimensional kinematic or detailed abundance information \citep{gilbert2022ApJ_trex, dey2023_desi, escala2023AJ_SPLASH_M31Inner}.}


Thereby, most resolved-star photometric studies of halos have used RGB stars along the minor axes of nearby galaxies as tracers of their accreted populations. Paired with inferences from cosmological simulations, these studies have revealed a diversity of stellar halo properties -- differences in metallicities, stellar masses, and color gradients -- as well as the long-term effects of these mergers on galaxy disks, star-formation histories, satellite populations, etc. Since stellar halo stars are mainly dominated by the contributions of the most (or a few of the most) massive progenitor(s) \citep{cooper2010, DSouzaBell2018a, monachesi2019, monzon2026arXiv260118954M}, resolved-star halo observations can make quantitative predictions about each galaxy's most important mergers and encode the effects of these mergers on the galaxies themselves. 

Wider-field photometric studies covering entire halos can also reveal observational signatures of mergers in the form of the aforementioned substructures such as shells, plumes, streams, loops, spurs, and overdensities. These features trace the distribution of baryonic and DM mass in the main galaxy \citep{merrifiendandkuijken1998, cron2016, bovy2016ApJ_streams_MWHaloshape, walder2025ApJ_probingDMhaloswstreams}, and also reflect interactions with satellite galaxies. However, to fully reconstruct a more detailed picture of a galaxy's merger history using these substructures --- including quantifying the timing of mergers and the orbit and angular momentum (AM) of an interacting satellite --- requires spectroscopy of their stellar populations, which provides information about their line-of-sight radial velocities and chemical compositions, \katya{as well as further enables the differentiation of accreted and \textit{in-situ} stars}.

\katya{Inferring the AM and orbit of accreted satellites is especially useful for observationally understanding phenomena such as how disks reorient due to mergers, which has already been seen in cosmological simulations. 
Multiple simulations show that 
a majority of \katya{accreted} halos also rotate in the same direction as the disk, and that 
disk tilting is a common response to satellite accretion, wherein galaxies reorient their disks to align their present-day AM with the orbital AM of their dominant progenitor (\citealt{gomez2017MNRAS_aurigadiskwarpreorient, dillamore2022MNRAS_ARTEMISdisktilt, bell2026OJAp_mergersAM}). Signatures of such reorientation can only be probed by multiobject halo star spectroscopic surveys.} 


One of the best examples of successful merger inferences from the combination of photometric and spectroscopic studies is our own Local Group (LG) -- extensive observations of the MW's stellar halo have uncovered that it is dominated by debris from the Gaia--Sausage--Enceladus (GSE) event, a smaller ($\sim10^9 M_\odot$), ancient (8-10 Gyr ago) merger, coinciding with the formation of its thick disk and globular cluster (GC) delivery \citep{bell2008, belokurov2018MNRAS_mwdiscandhalo, helmi2018, naidu2021ApJ_MWmajormergerH3survey}. Conversely, Andromeda (M31) shows a highly structured and asymmetric halo with prominent tidal debris, indicating that it likely had a recent dominant immigration event with a large ($10^{10} M_\odot$) galaxy $\sim 2-3$ Gyr ago, also causing disk thickening, a burst of star formation, and the delivery of a significant fraction of its satellites \citep{pandas, dsouzabell2018bm31, gilbert2018, Hammer2018, Williams2017,  DSouza2021}. \katya{Many studies have shown that several areas of M31's halo contain clear kinematic substructure that can be linked back to coherent photometric structures visible in its halo such as its Giant Stellar Stream, Northeast Shelf, and Western Shelf and building further evidence that most of its inner halo was built from a recent (1-2 Gyr) accretion \citep{gilbert2007ApJ_M31subGSS, fardal2012MNRAS_M31westernshelfspec,gilbert2009ApJ_SplashGSS,gilbert2019ApJ_abundanceGSS, escala2023AJ_SPLASH_M31Inner, dey2023_desi}. This substructure has also been used to constrain M31's gravitational potential and hence its dynamical mass at various radii and overall halo potential \citep{fardal2007MNRAS_m31streamshell, dey2023_desi}.}


Unfortunately, most other galaxies lack these kinds of important spectroscopic measurements --- such studies are sparse due to the prohibitively expensive nature of probing such faint and extended stellar populations. Discrete tracers such as planetary nebulae and GC populations are often used as proxies for detecting kinematic substructure in systems where resolved-star measurements are not feasible \citep{searlezinn1978ApJ_hierarchicalhaloformation,romanowsky2012,coccato2013MNRAS_accretioneventsinETGswPNEandGC,cortesi2013A&A_PnE_S0kinematics,kruijssen2019MNRAS_MWassemblythruGCs, tsakonas2025A&A_PNeM31, hartke2025FrASS_PNeaccretion, dolfi2021MNRAS_sluggs}, but this method has limitations such as relying on the difficult measurement of robust GC ages, inflated errors from sparser spatial and velocity sampling, and the challenge of differentiating between accreted and \textit{in-situ} clusters \citep{2015ApJ...800...13P_ngc3115haloandGCs,hughes2021MNRAS_GCsasmassdistributiontracers, kruijssen2019MNRAS_MWassemblythruGCs}.

Only recently have we been able to expand outside of our Local Group (LG) and begin to understand halo kinematics in more distant systems. \cite{beltrand2024_ngc4945} were able to make the first-ever spectroscopic measurement of the bulk stellar halo motion of a MW-mass galaxy outside the LG -- NGC 4945 -- with the Very Large Telescope (VLT)/MUSE integral field unit (IFU). Pointing at a 1' $\times$ 1' field at 34.6 kpc along the major axis of the galaxy for $\sim 8$ hours, they were able to resolve individual RGB and AGB stars. Though they were unable to measure individual velocities, by stacking spectra of halo RGB stars they found that NGC 4945 has a slightly counter-rotating halo with a rotational line-of-sight (LOS) velocity of $-$44 km/s (relative to NGC 4945's systemic velocity) and an upper limit of $42 \pm 22$ km/s velocity dispersion. Though this work was able to make the first-ever measurement of a bulk halo motion outside of the LG, it only covered a small area and was not sensitive enough to recover velocities of individual stars, necessitating the stacking of spectra to make the measurement and did not allow for examination of any possible substructure.

The advent of massively multiplexed multiobject, wide-field spectrographs is poised to ring in a new era of halo spectroscopy, enabling individual stellar velocity measurements of halo stars outside the LG. In this paper, we extend the work of 
\cite{beltrand2024_ngc4945} to carry out a stellar halo kinematic study of NGC 253. This nearly edge-on, starbursting MW-mass ($M_* = 5.5 \pm 1.4 \times 10^{10} M_\odot$; \citealt{Harmsen2017}) spiral is the brightest galaxy in the nearby Sculptor group/filament (D $\sim$ 3.5 Mpc; \citealt{radburn-smith2011ApJS_ghosts_ngc253dist}). Surveys of this system have identified that it hosts a highly flattened (b/a = $0.55_{-0.05}^{+0.04}$) stellar halo that is slightly more massive than the Large Magellanic Cloud ($4.5 \pm 1.9 \times 10^9 M_\odot$) with a prominent SW shell/shelf structure that can be traced up to 35 kpc from the center (see Fig. \ref{fig:map}) first seen in photographic imaging from 1982 and confirmed through resolved-star studies that point to an asymmetric halo \citep{malinandhadley1997, Greggio2014, Harmsen2017}. We note that \cite{Greggio2014} also identify another overdensity along the NW minor-axis at $\sim 30$ kpc, which is sometimes also referred to as a shelf \citep[e.g.,][]{harmsen2023MNRAS_agbrgb}. This paper focuses solely on the former SW structure, and this will henceforth be referred to as NGC 253's ``shell." This shell is theorized to be a signature of a tidal encounter/merger with a now almost phase-mixed companion \citep{beck1982A&A_ngc253shell, malinandhadley1997, davidge2010ApJ_ngc253disrupted, bailin2011, mutlu-pakdil2022ApJ_pisces}. 
The only work we know of that has aimed to quantify the timing of its dominant merger is \cite{harmsen2023MNRAS_agbrgb}. They used the ratio of AGB/RGB stars in GHOSTS minor-axis fields to estimate its star-formation history, finding that 90\% of star formation in NGC 253's halo was complete by $\sim$ 5-6 Gyr ago, but with considerable uncertainties in the calibration. In order to test whether we can detect signatures of dynamical youth in NGC 253's halo, we focus our observations on the region around its shell (centered 11 kpc in projected distance from the center of the galaxy), surveying a $\sim700\times$ larger area than \cite{beltrand2024_ngc4945}.

\urgent{Throughout the paper, we assume that NGC 253 has a heliocentric systemic velocity of 241 km/s \citep{HlavacekLarrondo2011} and is located at a distance of 3.5 Mpc ($\mu=27.70$) \citep{radburn-smith2011ApJS_ghosts_ngc253dist}, which corresponds to a scale of 1.02 kpc/arcmin.}

\section{Data}\label{sec:data}
\subsection{Target Selection}\label{sec:targetselection}

Because our goal was to measure individual stellar velocities, we chose to target the brightest stars available to us --- AGB stars. Younger and intermediate-age ($\sim$ 2.5 Gyr) stellar populations are expected to have a modest number of thermally pulsating AGBs brighter than the tip of the red giant branch (TRGB), and NGC 253 appears to host a population of these young AGB stars \citep{Greggio2014,harmsen2023MNRAS_agbrgb}. AGB stars are most prominent in the near-infrared, and so we select them using $Z$-band (depth $Z\sim24$ mag) and $J$-band (depth $J\sim23$ mag) photometry from a VISTA survey \citep{Greggio2014} of NGC 253. \moreurgent{All VISTA magnitudes are listed in the Vega system}. To assist with cleaning and sky fiber placement, we utilized $g$ and $r$-band imaging from the PISCES survey ($r \sim 26$; \citealt{sand2014ApJ_pisces}), and were able to test for contamination from background galaxies with deep resolved-star data from Hubble Space Telescope (HST) fields \citep{radburn-smith2011ApJS_ghosts_ngc253dist}. We corrected the VISTA and PISCES astrometry to the Gaia DR3 frame to an accuracy of $\sim0\farcs1$, sufficient for accurate Michigan/Magellan Fiber System (M2FS) fiber placement. \moreurgent{Contamination in any fiber from fainter stars is 
at least a factor of 6 fainter than our targets. At any rate, the sky surface brightness in Z-band is 600 times the galaxy's surface brightness, so contamination from fainter background stars is overwhelmed by uncertainties in sky subtraction (see Sec. \ref{subsec:skysubtraction}).} 

Candidates were selected from the VISTA catalog that lie outside projected major axis radius $R_{\textrm{maj}} > 15 \text{ kpc}$ (equivalent to a minor-axis distance of $R_{\textrm{min}} > 3.5 \text{ kpc}$) to minimize disk contamination. Comparison with HST photometry suggests that $\sim$85\% of AGB candidates (selected by color; Fig.\ \ref{fig:cmd}) from VISTA should be actual stars, while only $\sim15$\% are expected to be background galaxies. This region of color--magnitude space will also contain some foreground MW stars \citep{Greggio2014,harmsen2023MNRAS_agbrgb}. Given NGC 253's high Galactic latitude of $b = -87.96^\circ$, we do not expect significant MW foreground contribution. As seen in \cite{Greggio2014} who resolved NGC 253's stellar halo with VISTA, at the J$>$21.5 magnitude range in which we observe, NGC 253 stars dramatically outnumber foreground stars when compared to TRILEGAL simulations (see their Fig. 8). For stars for which we recovered velocities for (see Sec. \ref{sec:velmeasures}), we set an upper limit for the foreground contamination in our sample by measuring the percent of stellar velocities with respect to NGC 253 that are within $100$ km/s of -241 km/s. Only 13\% of our measured velocities are consistent with possibly being MW foreground. \urgent{We further discuss the possibility of MW contamination in \urgent{Sec. \ref{subsubsec:MWforeground}}}

Owing to the possibility of foreground and background contamination varying as a function of color, magnitude, and position, we avoid overweighting any given region of color--magnitude space or radial range, choosing instead to select a diverse subset of candidates with properties consistent with being AGB stars in NGC 253 by applying random weights, weighted by magnitude and equivalent projected major axis radius. We enforced strict exclusion zones around Shack-Hartmann (SH) stars, guide stars, and alignment stars. In particular, fibers were not placed within $10''$ of stars with $G < 21$, $30''$ for $G < 17$, or $60''$ for $G < 15$. We also choose stars to have no more than 1 neighbor in the final catalog within 14". Each fiber has a 1.2" aperture on sky, so we are well separated from any bright sources that could contaminate the flux of our targets. With these criteria in mind, our color/magnitude cuts for choosing stars are defined by the bounding box with \moreurgent{(Z-J, Z) vertices of [(0.6, 22.9),(1.2, 23.35),(1.5, 23.0),(0.8, 22.3)]. } We also choose a random subset of potential sky positions such that each sky fiber is between $30'' - 60''$ from a target star and, making sure that there are no PISCES (ground-based $g$ and $r$-band imaging; \citealt{sand2014ApJ_pisces}) sources brighter than $r<25.2$ within $2''$ of a sky position. Additionally, seven ``Telluric" stars (early F-type or bluer, $-0.8 < BP-RP < 0.4$) were observed. These stars, which lack intrinsic calcium infrared triplet (CaT) absorption, were used to characterize and remove telluric absorption features from the science spectra.

\begin{figure*}[!h]
    \centering
    \includegraphics[width=1\linewidth]{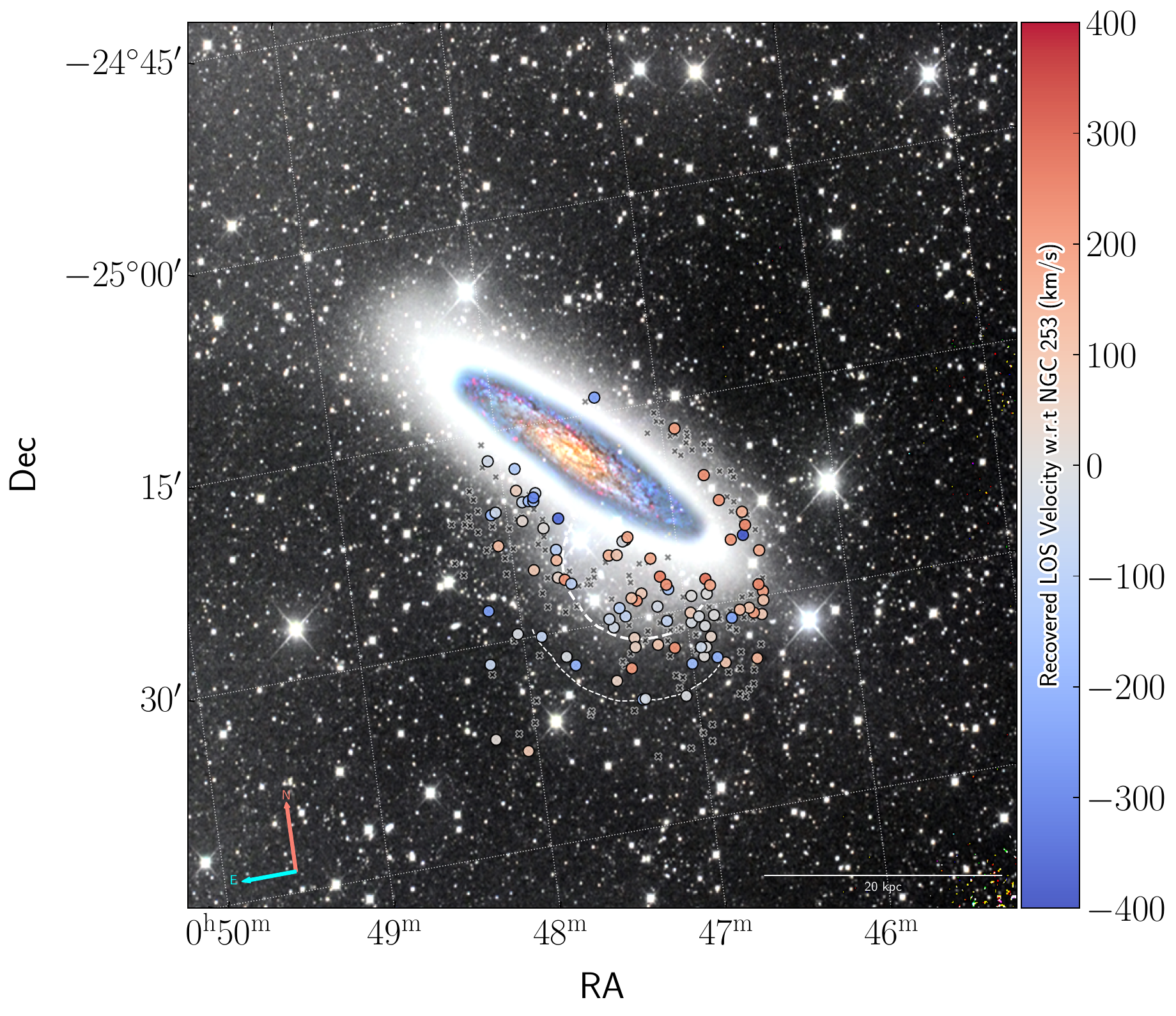}
    \caption{A spatial map of our target stars in NGC 253's halo. Circles are color coded by their recovered velocity with respect to NGC 253's systemic velocity, and black x's indicate that a velocity could not be recovered. \katya{The thicker and thinner white dashed lines indicate approximate contours of the SW shell as shown in Fig. 9 of \citealt{bailin2011}, corresponding to their dashed black line and their orange contour (corresponding to a density of $\sim5-5.7 \ \log{\rm{N}_{TRGB}/\rm{sq.deg}}$), respectively.} Original background image of the galaxy and integrated light halo by Michael Sidonio, which was then fed through Aladin to extract WCS and restacked by Katya Gozman. Despite similarities in color shades, we note that the colormap only applies to the recovered target stars surrounding NGC 253 and not to the colors of its disk, which are purely from false-color stacking of multifilter images.}

    \label{fig:map}
\end{figure*}

\begin{figure}[!h]
    \centering
    \includegraphics[width=1\linewidth]{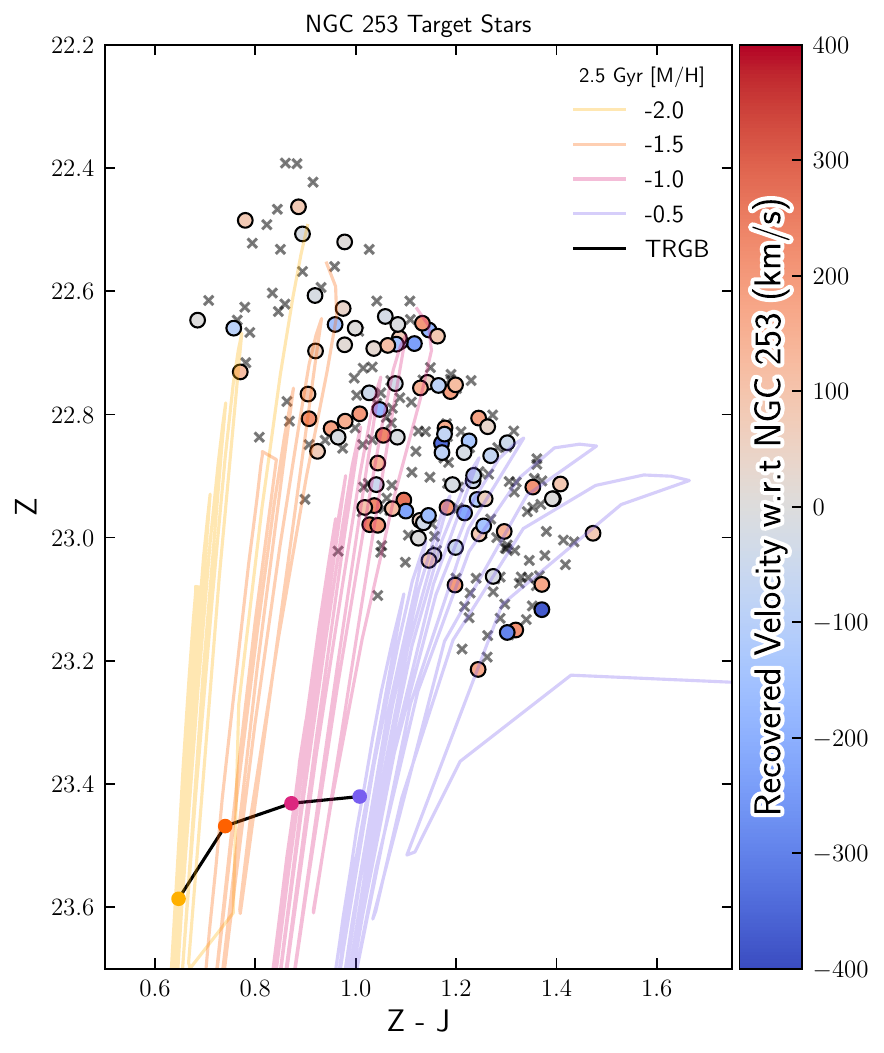}
    \caption{A NIR color-magnitude diagram of NGC 253 halo stars we target in this work. A selection of metal-poor, 2.5 Gyr PARSEC isochrones are shown as colored lines, and each star is either a circle color coded by its velocity with respect to NGC 253's systemic velocity if it was recovered, or a gray x if its velocity was not recovered. The black line with dots shows the location of the TRGB for each isochrone, assuming D=3.5 Mpc. Our target stars are well above the TRGB, placing them securely as AGB stars.}
    \label{fig:cmd}
\end{figure}

\subsection{Observations}\label{sec:observations}
We observed NGC 253's halo using the Michigan/Magellan Fiber System (M2FS; \citealt{mateo2012_m2fs}) on the Magellan/Clay telescope at Las Campanas Observatory in the 2022B, 2023B, and 2024B semesters. M2FS is a fiber-fed multiobject spectrograph capable of taking simultaneous observations of up to 256 objects at once. Due to poor seeing and weather conditions, our 2022B observations were not sufficiently deep to be scientifically useful; therefore, we only analyze data obtained from the 2023B and 2024B runs. M2FS has two twin spectrographs (called the ``B" and ``R" arms) which were set up identically for each run. We used M2FS in its low-resolution mode, with 2x2 binning, the 180 $\mu$m slit, and the R600 grating, yielding a plate scale of 0.25 arcsec/pixel. \urgent{Each fiber has a core diameter of 150 $\mu$m, which translates to 1.2" on sky.} The 2023B run utilized the BK7 filter, which yielded observations over a wavelength range of 690-869 nm with an effective resolution of $R\sim2000$. This range includes the Ca IR triplet, as well as the TiO and C bands in late-type stars. In 2024B, we used a previously unavailable longpass blocking filter that suppresses 2nd-order light, which can be an issue blueward of 7500\AA. This yielded a slightly larger wavelength range of 682.9-876.1 nm.
 
In 2023, out of M2FS's 256 total fibers, we dedicated 225 fibers to target stars, 24 to sky regions, and 7 telluric stars. One of the target fibers had very weak throughput, so we excluded it from analysis in this run and are left with 224 target fibers. After a preliminary analysis of our 2023B data, we decided to reallocate some of the target fibers to obtain more sky measurements. For the 2024B run, we discounted 3 dead or very weak fibers and were left with 208 target fibers, 38 sky fibers, and 7 telluric stars. \urgent{Together, we had 1 target that was unique to the 2024 run due to being a bad fiber in the 2023 run, and 17 targets unique to the 2023 run, 16 of which were reallocated to sky fibers in 2024 and 1 of which was a bad fiber in 2024.}


We also obtained ThAr, ThNe, and NeArHgXe lamps for wavelength calibration and twilight sky spectra for flat-fielding.
The observations and exposure times for the science targets are listed in Table \ref{tab:observations}.

\begin{table*}
\resizebox{\textwidth}{!}{%
\begin{tabular}{l|l|l|l|ll}
\multicolumn{1}{c|}{Semester} & 
 \multicolumn{1}{c|}{Night} & \multicolumn{1}{c|}{Num. Targets} & \multicolumn{1}{c|}{\# Exposures} & \multicolumn{2}{c}{Exposure Times (s)} \\ \hline
 &  &  & &  \multicolumn{1}{c|}{B} & \multicolumn{1}{c}{R}\\ \hline
2023B & \urgent{UT20230916} & 224 & 6 & \multicolumn{1}{l|}{2700, 2400, 2400, 2400, *2600, 2700} & 2700, 2400, 2400, 2400, *2400, 2700 \\
2023B & \urgent{UT20230917} & 224 & 6 & \multicolumn{1}{l|}{2700, 2400, 2400, 2700, 2700, 3600} & 2700, 2400, 2400, 2700, 2700, 3600 \\
2024B & \urgent{UT20241002} & 208 & 8 & \multicolumn{1}{l|}{3600, 3600, 3600, 3600, 3600, 3600, 3600, 1800} & 3600, 3600, 3600, 3600, 3600, 3600, 3600, 1800 \\
2024B & \urgent{UT20241003} & 208 & 9 & \multicolumn{1}{l|}{3600, 3600, 3600, 3600, 3600, 3600, 3600, 3600, 1800} & 3600, 3600, 3600, 3600, 3600, 3600, 3600, 3600, 1800
\end{tabular}%
}
\caption{Exposure Times for 2023B and 2024B observations. Total exposure time is 24.8 hours. Exposure times marked with a * are the only ones that were different between the B and R sides of the spectrograph in a given exposure.}\label{tab:observations}
\end{table*}

\subsection{Data Reduction}\label{sec:reduction}

We reduced our data using standard \texttt{IRAF} routines.  Each spectrograph has a CCD with images from four amplifiers used to read out the CCD, which have slightly different noise and gain. We start by trimming the overscan region and subtracting the bias level from each individual amplifier image, then rotating and stitching together the four images and performing dark current correction. We then remove cosmic rays using \texttt{LA Cosmic} \citep{vandokkum2001PASP_lacosmic}. The pipeline then extracts each fiber spectrum in each image and wavelength-calibrates each spectrum using the arc lamps by fitting a polynomial to the Ne, Xe, Hg, and Ar emission lines. All spectra were also normalized by fitting a polynomial to the twilight images, dividing by the average of those so the fitted polynomial image mean value is 1.0, then dividing by that image to remove the fiber-to-fiber variations. We also compute the variance of each spectrum, accounting for Poisson, readout, rms, and background noise. 



After this, we performed throughput correction (TC) separately for the B and R sides of the spectrograph to normalize the effects of differing fiber throughputs (from fiber losses, imperfect fiber alignment, etc.). We divide each fiber in a given B/R twilight exposure by the median of all fibers in that exposure, giving us the TC for each night, side, and fiber. We then find the median TC for each of our $\sim$128 fibers on each side\footnote{We exclude 1(3) weak/dead fibers from the 2023(2024) data from our analysis}, apply a 201-pixel-wide median filter, and divide the flux from each corresponding fiber by the filtered TC and the variance by the square of the filtered TC.


Due to longer exposure times for some of our fields, we find that \texttt{LA Cosmic} misses some cosmic rays (CRs) that are present in the data. Therefore, we identify and correct these by hand for the 2024B data, which had the longest exposure times (up to 1 hour per field). We visually inspect each spectrum by hand and note fibers that contain CRs. For each fiber in a given exposure, we find all the other fibers of the same type not affected by CRs (target, sky, or telluric) in that exposure and calculate a median spectrum of these fibers. We then subtract that from the CR-affected spectrum and find the index of the maximum value, which is the location of the cosmic ray. To correct this, the values of 3 pixels on either side of this index are replaced with the values of the median spectrum calculated earlier. We repeat this procedure if necessary for fibers containing more than 1 CR. For the variance spectra, we calculate the ratio between the variance and the data, subtract off a 55-pixel median filter of this ratio, and find the indices where the median-subtracted ratio is greater than the median value of the ratio + 30$\times${\texttt{mad\_std}}(ratio), where we use \texttt{mad\_std} to refer to the median absolute deviation as calculated by \texttt{astropy.stats.mad\_std} for the remainder of this work. We then set the values of 4 pixels on either side of each identified peak to the (median-filtered ratio+the data) in these 8 pixels $\times$ median filter of the ratio.

\section{Methods}
\subsection{Sky Subtraction}
\label{subsec:skysubtraction}
Given that the AGB stars we target are, on-average, Z$\sim$22.9, we require robust sky subtraction to isolate the signal of each star from prominent sky emission lines. At our red to near-IR wavelengths, the sky emission is mostly dominated by varying groups of OH bands between $\sim 7000-9000$\AA{} and O$_2$ bands around $\sim 8600$\AA, which can show strong variability in intensity over a night and from place to place across our field of view (FOV). Though each of our exposures contains sky fibers that can be used for a basic sky subtraction, the combination of the complex behavior of airglow lines and the low signal-to-noise (S/N) of our spectra necessitated a more nuanced, two-step approach.
\subsubsection{Skycorr}

The first step was to apply the program \texttt{Skycorr} \citep{skycorr2014} to our data and sky fibers, which performs a physically motivated scaling of airglow lines to fit sky lines in target spectra. We present an overview here, though we refer the reader to the original \cite{skycorr2014} paper for a detailed discussion of the methodology. \texttt{Skycorr} takes in an input science spectrum and a reference sky spectrum. It then identifies pixels belonging to either sky lines or continuum and finds groups of related sky lines in the sky spectrum using an airglow model from \cite{noll2012A&A_airglowmodel} for the specific date and time the spectra were taken. It also detects any deviations in wavelength calibration between the sky and target spectra and rebins the spectra, after which it scales the reference sky lines to fit the sky lines in the target spectrum, and subtracts the scaled sky lines and sky continuum from the target spectrum to get a resulting sky-subtracted target spectrum. \katya{\texttt{Skycorr} also outputs an error (variance) array for each spectrum that we use for the remainder of our analysis unless stated otherwise as the noise for any given exposure.}

We experimented with different combinations of sky fibers to feed into \texttt{Skycorr,} such as giving it a median of all sky fibers in a given exposure or finding the closest sky fiber to each target, but since we wanted to get spectra that were as free of sky-subtraction residuals as possible, for each target we iterate through each sky fiber in a given exposure, run \texttt{Skycorr}, and choose the resulting sky-subtracted spectrum that has the smallest residuals. This is quantified by continuum-subtracting each resulting sky-subtracted spectrum, weighting it by its noise spectrum, and finding the standard deviation of flux in sky-free regions, as identified by \texttt{Skycorr}. We kept the result of the sky fiber that yielded the smallest standard deviation and used the corresponding sky-subtracted spectrum for the rest of our analysis. When compared to using the nearest sky fiber, this technique decreased the median ratio between the \texttt{mad\_std} of the most prominent sky regions to continuum regions by $11\%$  and decreased the median ratio between the rms of these regions by $14\%$.

\katya{The sky-free regions discussed in the previous paragraph are identified as such. The outputs from \texttt{Skycorr} contain an output sky-subtracted spectrum and an error array. Each pixel is also tagged as being continuum (class = 0) or part of a sky line (class $\neq$ 0), which has minor variation exposure-to-exposure. To make a consistent classification, we choose to tag a pixel as continuum if 90\% of all spectra have that pixel classified as continuum by \texttt{Skycorr}, and tag it as sky otherwise. We adjust the classification by inspecting some spectra by hand and manually tagging pixels where we see significant emission that was misidentified by \texttt{Skycorr}.}


\subsubsection{Principal Component Analysis}

Once we ran \texttt{Skycorr} on all target, telluric, and sky spectra, we noticed many spectra that still had considerable systematic sky-subtraction residuals (see middle panel of Figure \ref{fig:reduction}). This is a common problem in multi-fiber data, especially at redder wavelengths where airglow lines are abundant and most prominent. We attempt to correct these residuals using a method based on principal component analysis (PCA), mostly following \cite{wild_hewitt_sdss_pca_2005MNRAS.358.1083W}'s methodology. This technique has been used to rectify residuals in a number of fiber surveys and instruments, including LAMOST \citep{bai2017RAA_pca_lamost}, VLT/MUSE \citep{zap2016MNRAS_soto}, Anglo-Australian Telescope (AAT)/AAOmega \citep{sharp_parkinson_2010MNRAS_aat_pca}, and Sloan Digital Sky Survey \citep[SDSS;][]{wild_hewitt_sdss_pca_2005MNRAS.358.1083W}. The general goal of PCA is to create a set of orthogonal components that represent the systematic residuals that come about from sky subtraction. We can reconstruct our sky residuals with a linear combination of these components, apply them to our sky-subtracted targets to fit these residuals, and then subtract them from our target spectra. We outline the details of our method below.

We first continuum-subtract all of our spectra by subtracting a 55-pixel median filter of either target, sky, or telluric spectra, depending on which spectrum we are continuum-correcting. \katya{Since PCA finds components with the greatest variance, it is important to remove nonphysical outliers because we want to isolate only sky-subtraction residuals to eliminate from our spectra.} Therefore, we remove any outliers in the resulting spectra by finding pixels in each spectrum that are more than 10 times the \texttt{mad\_std} away from the median value of the spectrum and setting an 8-pixel-wide window centered around the outlier pixel to values of the median of all the spectra of its corresponding type (target, sky, or telluric) in that region. 

Once the spectra are continuum subtracted and outliers are removed, we then proceed to find the optimal number of components needed for PCA by using our sky-subtracted sky spectra (referred to as just ``sky spectra" here), as these should not contain any signal, only the sky-subtraction residuals that we want to subdue. We mask the continuum pixels identified by the previous step and noise-weight each sky spectrum and perform a PCA fit and reconstruction on these noise-weighted masked sky spectra, which is then subtracted from the nonmasked sky spectra to figure out the specific number of components to use in PCA reconstruction. We follow the logic of \cite{wild_hewitt_sdss_pca_2005MNRAS.358.1083W} by calculating the ratio of the \texttt{mad\_std} in regions of sky and nonsky pixels and finding the number of components that will yield a mean ratio equal to 1, such that we are using enough components to capture sky residuals without artificially removing noise below the Poisson limit. This yielded 6 components for the 2024B data and 10 components for the 2023B data, which we then use to fit the sky spectra principal components, project the corresponding masked target spectra onto these principal components and transform these back to the original space to get the residuals that we then subtract from the target spectra to get the final corrected sky-subtracted target spectrum. This process is also done for the sky and telluric spectra as well. This decreased the median ratio between the \texttt{mad\_std} of the most prominent sky regions and continuum regions by another 16\% from the ratio measured after performing \texttt{Skycorr}. We also note that if using the nearest sky fiber, one would require 36 components instead of 6, highlighting that using the sky fiber that minimizes the standard deviation of flux in sky-free regions in sky-subtracted spectra is effective at capturing a lot of the systematic sky-subtraction residuals.

\subsection{Spectral coaddition and Stacking}
\label{subsec:spectralcoadditionstacking}

After sky-subtracting, we perform routine telluric correction by dividing each spectrum by a smoothed median of our telluric spectra. With this last correction, we now combine individual exposures for each star into stacks. Because the 2023 and 2024 data have slightly different wavelength ranges and wavelength binning, we rebin the 2023 data to the same wavelength grid as our 2024 data using the \texttt{FluxConservingResampler} from Python's \texttt{specutils} package. \katya{This yields a binning of 1$\mathrm{\AA}$ per pixel.}

\katya{While the variance arrays from \texttt{Skycorr} are our most reasonable estimate of the uncertainty in the spectra, given how dominant sky noise is in some wavelength regions, we wish to quantify the importance of the large differences in noise from sky lines and continuum regions by exploring two additional cases for velocity fitting. For the first case, we adopt a wavelength-independent noise term. For the second case, we inflate the noise in regions with emission substantially in excess of the continuum by an inflation factor ($IF$). The inflation factor is computed as follows. We find the standard deviation per wavelength of all the individual target and sky fiber arrays and square it to get a variance spectrum. We then subtract the typical variance in low-variance regions between the sky lines from this, clip any negative results to zero, and take the square root of the remainder. Multiplying this by a factor of 3 yields the inflation factor, $IF$. The factor of 3 was chosen through extensive experimentation with velocity recovery. As discussed later in Sec \ref{sec:Marz}, we demand that the recovered velocities for a given spectrum are similar between these three different noise cases (uniform, default, or inflated), giving confidence that the recovered velocities are not unduly sensitive to our assumptions about noise.}

We combine single exposure spectra and also combine all star spectra in two different ways --- computing the median or weighted average. We add the inflation factor in quadrature to the noise from medianing or weighted averaging our individual exposures if relevant.

\begin{itemize}
    \item Medianing together single exposures to get star spectra and then medianing those to get a total median spectrum of the halo. If the default noise from \texttt{Skycorr} in each individual spectrum is $\sigma$, then the noise of the median was estimated as $<\sigma>\frac{\sqrt{\pi/2}}{\sqrt{N}}$, where N=number of spectra contributing to the stack. \katya{The $\sqrt{\pi/2}$ is used because we are finding the error in the median rather than the mean, and this assumes a normally distributed population \citep{Williams2001}. For the inflated noise, this becomes $\sqrt{\left(<\sigma>\frac{\sqrt{\pi/2}}{\sqrt{N}}\right)^2 + \left(\frac{IF}{\sqrt{N}}\right)^2}$}
    \item A weighted average by weighting the flux F of each spectrum by the noise: $\frac{\Sigma\left(\text{F}/\sigma^2\right)}{\Sigma(1/\sigma^2)}$. The noise is estimated as $\sqrt{\frac{1}{\Sigma(1/\text{noise}^2)}}$ and for the inflated errors, $\sqrt{\left(\sqrt{\frac{1}{\Sigma(1/\text{noise}^2)}}\right)^2 + \left(\frac{IF}{\sqrt{N}}\right)^2}$
\end{itemize}

\begin{figure*}
    \centering
    \includegraphics[width=0.99\linewidth]{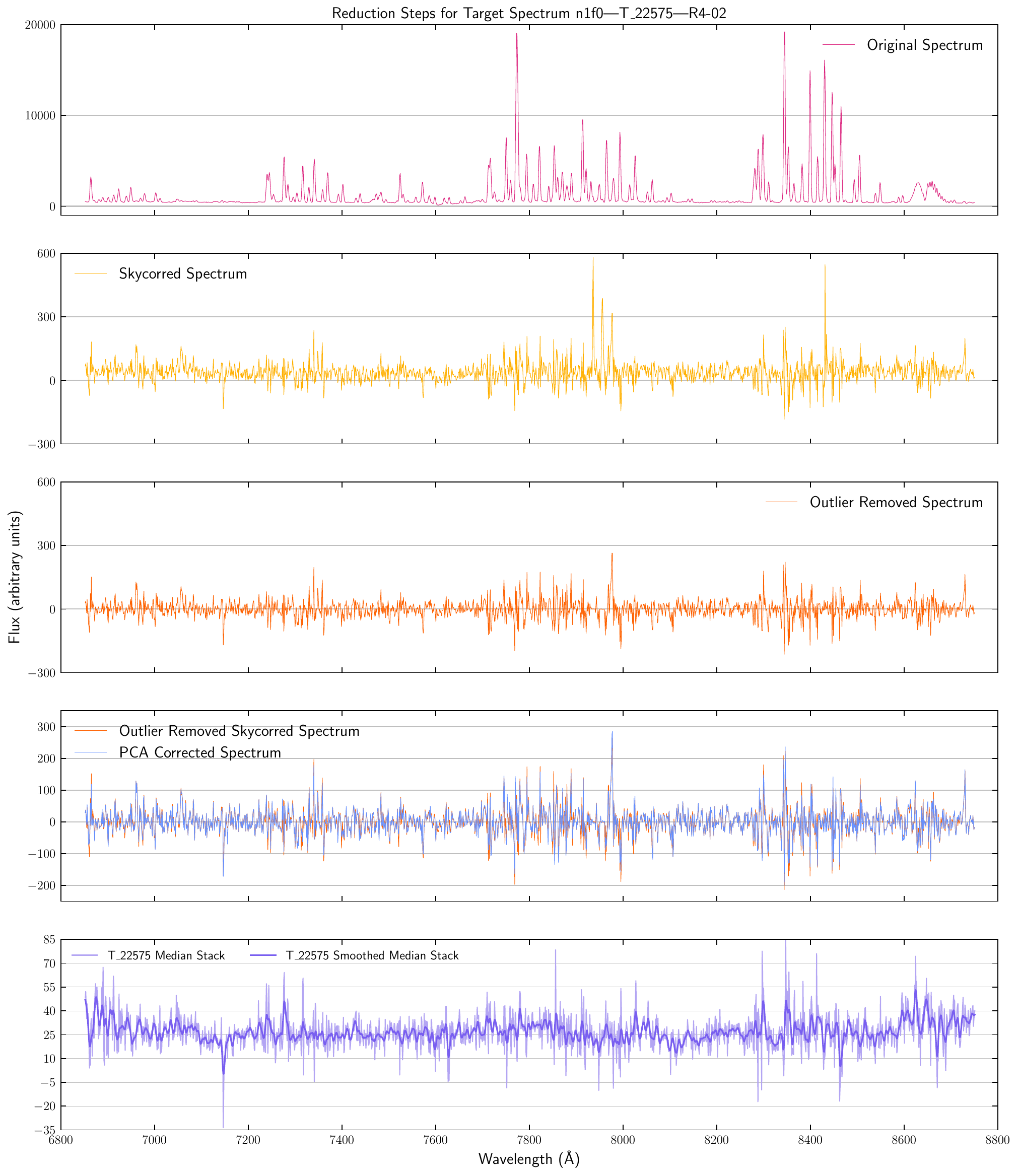}
    \caption{An example of the reduction steps for a single star spectrum. The top panel shows a reduced, normalized, wavelength-calibrated spectrum with evident sky emission lines. The second panel shows the spectrum after it has been sky-subtracted through \texttt{Skycorr}. The middle panel shows the removal of outliers from the \texttt{Skycorr}ed spectrum and the fourth shows that spectrum after correction by PCA. The bottom panel shows the median stack of all the individual, sky-subtracted and telluric-corrected, exposures of the single target, with the darker purple showing the stack after being smoothed by a Gaussian filter of kernel size = 2.}
    \label{fig:reduction}
\end{figure*}

A series of plots showing the steps of the reduction process is illustrated in Fig. \ref{fig:reduction}.

\katya{Our uncertainty in the noise and dominance of sky residuals lead to spectra that do not have a well-defined continuum, precluding us from citing precise values for the S/N for each individual spectrum or measuring stellar metallicities. Instead, we derive an approximation for the average S/N of a stacked spectrum for an individual star. We do this by stacking together all 89 spectra for which we successfully recover a velocity (see Sec. \ref{sec:Marz} for details), finding the median S/N in regions not dominated by sky lines, and dividing by $\sqrt{89}$. We purposefully stack \textit{unshifted} spectra, as we want to keep them in the observed frame of reference since this is the frame in which we see and subtract sky lines. For the noise, we use the noninflated variances as discussed in Sec. \ref{subsec:spectralcoadditionstacking}. This yields an average S/N $\sim 2.1/\mathrm{\AA}$ per star.}


\subsection{Measuring velocities with \texttt{Marz}}
\label{sec:Marz}

We now detail how we attempt to measure the velocities of individual star stacked spectra \katya{and identify stars that likely belong to NGC 253}. We use the software \marz\ to perform cross correlation on each of our target spectra with a set of stellar templates. We refer readers to the original \marz\ paper \citep{hinton2016A&C_marz} for complete details on the software, but we summarize it briefly here. Originally written for the redshifting of spectra from the Australian Dark Energy Survey with the 2dF spectrograph, it has since been generalized to allow for the velocity fitting from any spectrograph. It uses the \texttt{autoz} cross-correlation algorithm to compare an input spectrum against a set of (in our case, stellar) templates and finds the peaks in the correlation functions, with the best-fit velocity resulting from the highest peak found between all of the templates. \urgent{An advantage to \marz\ is that the algorithm includes a set of preprocessing steps to remove bad pixels and CRs. It also performs error adjustment, which accounts for underestimation of the variance spectrum due to poor sky subtraction.}

As \marz\ only comes with 12 templates that model either galaxies, quasars, or main sequence stars which do not contain the stellar features we expect in AGB stars, we make our own set of stellar templates to use with the package. We use the POLLUX database \citep{palacios2010A&A_pollux} to query for a set of synthetic spectra from the \urgent{Archéologie avec Matisse Basée sur les aRchives de l’ESO (AMBRE)} grid of high-resolution stellar spectra \citep{delaverny2012A&A_ambrespectra}, which uses \urgent{Model Atmospheres with a Radiative and Convective Scheme (MARCS)} model atmospheres. We choose stellar parameters matching those of early M giants -- akin to the spectral types expected of AGB stars -- with $3500 \leq T_{\rm eff}/{\rm K} \leq 3700$, $0 \leq \log g \leq 1.5$, $-1.5 \leq \text{[Fe/H]} \leq -0.5$ and $0 \leq \alpha \leq 0.8$, which yielded 144 matches. We degrade these extremely high resolution spectra down by convolving them with our average line-spread function (a Gaussian of width 1.65\AA) and rebinning them to match our wavelength grid.

As our individual spectra have such a low S/N and thus no outstanding spectral features that we can use to tell by eye how good a cross correlation fit is, it was imperative that we understand how to quantify how well \marz\ can recover individual velocities and understand the error in recovered velocity over a range of S/N. Additionally, \marz\ always provides a best-fit velocity, but due to our uncertainty in our noise estimates, traditional goodness-of-fit measures like $\chi^2$ are not useful for distinguishing between high- and poor-quality fits, \eric{requiring selection criteria that rely on the recovered velocities only. In order to develop these criteria, }
we created a set of mock spectra that could mimic our observations by adding different combinations of template spectra and sky-subtracted sky spectra (to add in realistic sky residuals and degrade the signal). We fit these spectra with \marz, varying different spectral and fitting parameters, and evaluated the accuracy of each fit (the offset between the fit and input velocity), the spread in recovered velocities as a function of different metrics, and the $F_1$ score for each combination of metrics. The details of this mock analysis are discussed in Appendix \ref{sec:appendix}. These experiments showed that the stability of recovered velocity is influenced by changes in assumed noise and velocity range used for \marz\ fitting, and to whether we input a median stack or weighted-average stacked spectrum. \eric{We use these mocks to} 
determine a set of recovery criteria based on these different fitting choices that give us the best tradeoff between the number of true positives, false negatives, and false positives that we recover.

Our recovery criteria require combining the velocities from different \marz\ fits with 18 different combinations of spectral coaddition, noise, and \marz\ maximum fitting values:



\begin{enumerate}
    \item Each spectrum is the result of coadding with a median or a weighted average as referenced in Sec. \ref{subsec:spectralcoadditionstacking}.
    \item The noise is set to either the standard deviation spectrum of that fiber from \texttt{Skycorr}, the standard deviation spectrum with artificially inflated errors as done in Sec. \ref{subsec:spectralcoadditionstacking}, or a uniform standard deviation value of 100 across all wavelengths.
    \item \marz\ does not let the user specify a starting velocity, but it does let the user set a velocity range over which to execute the cross correlation. We experiment with different velocity ranges to act as a robustness check -- if the best-fit velocity does not drastically change with varying velocity ranges, we consider that a signature of a more robust fit. We limit the fits to ranges of $\pm$[1800, 3000, 6000] km/s (much larger than the expected stellar velocities of our stars). 
\end{enumerate}

Our criteria for robust velocity recovery combine the output velocities for these 18 values per target (see also Appendix \ref{sec:appendix}), by demanding that they satisfy a combination of conditions. We demand that the 18 velocity values satisfy one or more of the following three conditions A)-C): 
\begin{itemize}
    \item \urgent{(A) all but two\footnote{\katya{As discussed in Appendix \ref{sec:appendix}, based on our extensive experimentation using realistic mock datasets for which the true velocity is known, using all but two spectra gave us a good balance between recovering correct velocities that may have had 1-2 outlier measurements but are otherwise robust while not letting in too many outliers with incorrectly determined velocities.}} coadd weighted-averaged spectra within a group had recovered velocities that were within $\pm$50 km/s of the median recovered velocity of all coadd weighted-averaged spectra within a group OR}
    \item \urgent{(B) all but two coadd median spectra within a group had recovered velocities that were within $\pm$50 km/s of the median recovered velocity of all coadd median spectra within a group OR}
    \item \urgent{(C) the absolute value of the difference between the median recovered velocity of the average-weighted spectra and the median spectra was $\leq 100$ km/s.}
    \item \urgent{(D) In addition, we select as stars  only targets with recovered velocities that were within $\pm400$ km/s from NGC 253's systemic velocity of 241 km/s. Criterion D) allows for a large velocity dispersion, and anything beyond 400 km/s is likely to be either an unbound star or nonphysical measurement. We remind the reader that we are not limiting \marz\ to fit only velocities in this range, but rather just doing a simple cut on the velocities it returns.}
    
\end{itemize}   
\urgent{66 stars pass criteria (A) OR (B), 119 pass criterion (C); for the 135 stars that pass (A) OR (B) OR (C), 89 of them also pass criterion (D).} From our mock simulations, we saw that discarding stars that satisfied one or more of criteria (A)-(C) with velocities outside the $\pm400$ km/s range (thereby failing criterion (D) managed to exclude most spurious measurements. If a spectrum meets these criteria, that star is deemed recovered, else we consider the spectrum to be too noisy to recover a robust velocity.

Changing the recovery criteria to other sets that are deemed good (e.g., high $F_1$ scores and higher percentages of total spectra deemed recovered) with our mocks does not significantly alter our results (e.g. replacing criteria (A) and (B) with whether the \texttt{mad\_std} of the velocities within a group is $\leq 50$ km/s recovers $\sim3$\% fewer stars but with the same median velocity and velocity dispersion as with our chosen criteria, within error).




\section{Velocity Measurements}
\label{sec:velmeasures}
\subsection{Individual Halo Star Velocities}

The above criteria yield 
89 stars (40\% of our total sample) that we choose to focus on for our velocity measurements and further analysis. \katya{Because we believe our variance spectra to be underestimated}, we take the error on the velocity of a star from our mock spectra analysis. Instead of trying to derive an uncertainty for each star independently, we adopt the average of the 16th and 84th percentiles of the velocity offset in our outlier-free recovered mocks \katya{(48.1 km/s and 27.7 km/s, respectively; see Appendix \ref{sec:appendix}) for a detailed discussion} as the uncertainty on each of our stars. This yields an uncertainty of 40 km/s on each stellar velocity. 

\katya{The CMD presented in Fig. \ref{fig:cmd} has been color coded by the recovered velocity of each star with respect to NGC 253, if applicable. There appears to be no correlation between the magnitude of a star and whether or not we were able to recover a velocity --- in fact, many of the brightest sources do not have measured velocities. We suggest two main reasons for this. One is simply that our chosen targets are slightly contaminated by nonstellar sources --- as discussed in Section \ref{sec:observations}, $\sim15$\% of AGB candidates in VISTA may be background galaxies, and as such would not have spectral features that match our AGB-specific stellar templates. The other reason is that sky-subtraction residuals dominate whether a velocity recovery is possible. Indeed, our mock observations were made to simulate stars spanning the magnitude range we observe with realistic sky-subtraction residuals, and we only recover velocities for $\sim 50\%$ of them, indicating that sky subtraction is a major bottleneck that allows or prevents a robust velocity measurement. This mock recovery fraction is quite similar to the recovery fraction of our data (40\%), making it a reasonable recovery fraction given the importance of sky residuals for velocity recovery.} \urgent{We list the details of each targeted star and its measurements (where applicable) in Table \ref{tab:stars}.}

\begin{deluxetable*}{lccccccccc}
\tabletypesize{\scriptsize}
\tablewidth{0pt}
\tablecaption{NGC 253 Target Star Properties\label{tab:target_properties}}
\tablehead{
\colhead{ID} &
\colhead{R.A.} &
\colhead{Dec.} &
\colhead{R.A.} &
\colhead{Dec.} &
\colhead{$J$} &
\colhead{$Z$} &
\colhead{Exposure Time} &
\colhead{Velocity} &
\colhead{Label} \\
\colhead{} &
\colhead{(h:m:s)} &
\colhead{(d:m:s)} &
\colhead{(deg)} &
\colhead{(deg)} &
\colhead{(mag)} &
\colhead{(mag)} &
\colhead{(s)} &
\colhead{(\ensuremath{\mathrm{km\,s^{-1}}})} &
\colhead{}
}
\startdata
T20800 & 00:47:28.2035 & $-25$:34:10.4677 & 11.867515 & $-25.569574$ & 21.605 & 22.748 & 89100 & 48 & 0 \\
T21295 & 00:47:24.4425 & $-25$:33:11.2033 & 11.851844 & $-25.553112$ & 21.626 & 23.044 & 89300 & \nodata & \nodata \\
T19475 & 00:47:25.1736 & $-25$:32:53.5908 & 11.854890 & $-25.548220$ & 21.545 & 22.735 & 89300 & \nodata & \nodata \\
T22343 & 00:47:26.7464 & $-25$:30:21.8216 & 11.861443 & $-25.506062$ & 21.671 & 22.750 & 89100 & $-34$ & 0 \\
T22438 & 00:47:27.9375 & $-25$:29:43.5823 & 11.866406 & $-25.495440$ & 21.674 & 22.908 & 89100 & $-46$ & 0 \\
T18831 & 00:47:26.1736 & $-25$:28:20.5943 & 11.859057 & $-25.472387$ & 21.515 & 22.745 & 89100 & \nodata & \nodata \\
T22320 & 00:47:39.3874 & $-25$:36:14.4415 & 11.914114 & $-25.604012$ & 21.669 & 22.492 & 89100 & \nodata & \nodata \\
T19973 & 00:47:38.8204 & $-25$:35:59.8778 & 11.911752 & $-25.599966$ & 21.568 & 22.687 & 89100 & \nodata & \nodata \\
T18874 & 00:47:42.3175 & $-25$:32:33.2181 & 11.926323 & $-25.542561$ & 21.517 & 22.663 & 89100 & $-181$ & 0 \\
T20968 & 00:47:45.2325 & $-25$:31:50.3027 & 11.938469 & $-25.530640$ & 21.613 & 22.507 & 89100 & $-14$ & 0
\enddata
\tablecomments{\urgent{\nodata\ indicates that no measurement or classification is available. All stars with velocity measurements have an assumed measurement uncertainty of $\pm40$ km/s from mock spectra experiments as discussed in Appendix \ref{sec:appendix}. J and Z band magnitudes are derived from the VISTA survey of NGC 253 \citep{Greggio2014}. The ``Label'' column indicates the label assigned to each star with a measured velocity by Spectral Clustering; we interpret 0 to be the accreted shell region, 1 to be the \textit{in-situ} disk, and 2 to be the accreted halo. We caution that these labels only reflect the output from SC; ascertaining the true physical nature of each individual star requires deeper and higher-resolution data. This table is truncated; a full machine-readable table is available online.}}
\label{tab:stars}
\end{deluxetable*}

\subsection{Halo star stacking}

\begin{figure*}
    \centering
    \includegraphics[width=0.99\textwidth]{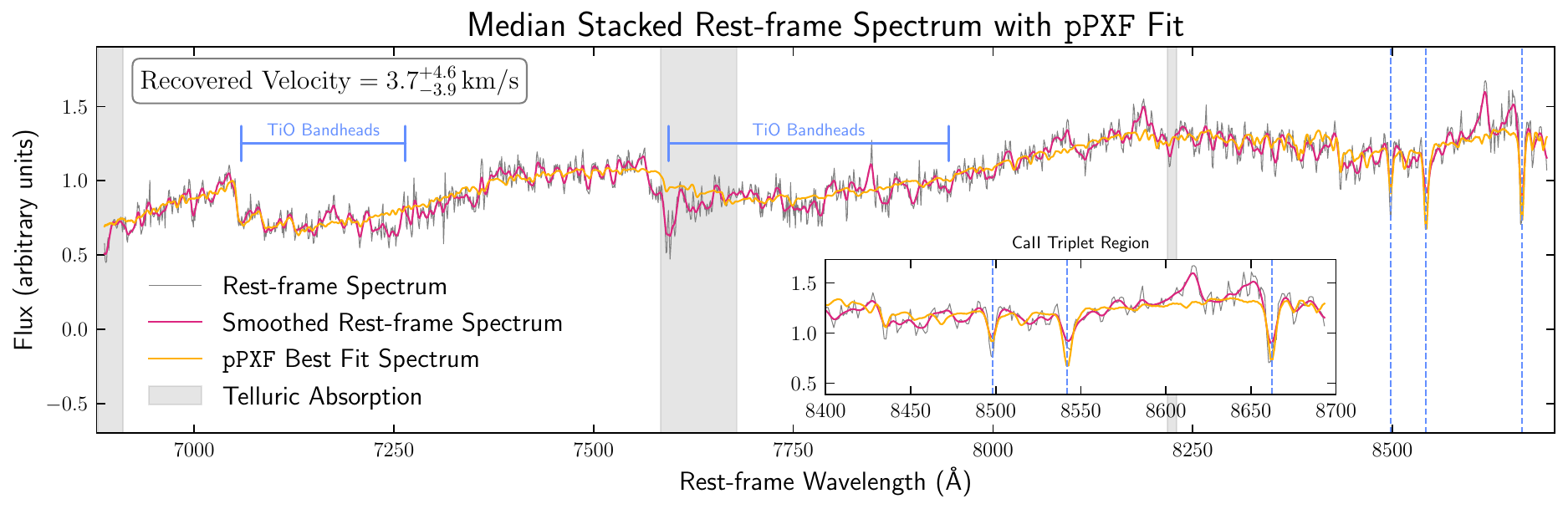}
    \caption{The stacked spectrum of our recovered stellar sample where each individual median coadd spectrum has been shifted back to the rest frame according to the recovered velocity. The gray spectrum is the raw coadd, the pink spectrum is the raw coadd smoothed with a Gaussian filter of kernel size = 2, and yellow spectrum is the \ppxf\ fit to the stack. The CaT lines are clearly visible, as are some TiO molecular bands. gray bands show regions of significant telluric absorption. \texttt{pPXF} recovers a velocity that is synonymous with 0 km/s within error, signifying that the stack is indeed in the rest frame and the velocities we recover with \marz\ are accurate.}
    \label{fig:ppxf_fit}
\end{figure*}

As a test of our individually measured velocities, we take the 89 stars in our recovered group, shift their spectra back to the rest frame in accordance to their recovered velocities, and median stack the resulting spectra. The resulting spectrum is shown in Fig. \ref{fig:ppxf_fit} as a gray line, with a pink line showing the same spectrum but smoothed by a Gaussian kernel with a width of 2. Initially indistinguishable in individual stellar spectra, we now see clear calcium triplet (CaT) infrared absorption lines, as seen in the inset panel of Fig. \ref{fig:ppxf_fit}, and a forest of molecular titanium oxide (TiO) bandheads. These two features are signatures seen in M-type, cooler stars, which is what we expect to see in our targeted AGB stars. 

To check whether the deredshifted stacked spectrum actually has a velocity of zero, we used the Python package \ppxf\ (penalized PiXel Fitting; \citealt{capellari_emsellem_2004PASP_original, capellari_2017MNRAS_ppxf, ppxf_Cappellari2023}), which recovers a line-of-sight velocity distribution (LOSVD) through maximum penalized likelihood full-spectrum fitting. \ppxf\ models the observed spectrum by fitting a linear combination of differently weighted templates provided by the user using Gauss-Hermite expansions. Though commonly used to fit superimposed stellar populations, it has been successfully used for fitting individual stellar spectra \citep[e.g.,][]{beltrand2024_ngc4945}. Using a set of 30 templates randomly chosen from the set of AMBRE templates used in \marz \ fitting, we fit the stacked spectrum with fourth degree additive and multiplicative polynomials (the \texttt{mdeg} and \texttt{deg} keywords) \katya{using a median error array computed from the inflated errors as discussed in Sec. \ref{subsec:spectralcoadditionstacking}.}. The resulting fit from \ppxf\ is shown as a yellow line in Fig. \ref{fig:ppxf_fit}. We recovered a \ppxf\ velocity of $3.7^{+4.6}_{-3.9}$ km/s with a negligible velocity dispersion. The error is derived from bootstrapping the spectra that are stacked, fitting each resulting stack, \katya{and reporting the 16th/84th percentiles of the measured velocities.} The \ppxf\ velocity indicates that within error, our spectrum is truly at rest frame. 



We also experimented with using \ppxf\ to do the individual velocity fits instead of \marz, but ultimately decided on \marz\ as it yielded $\sim3$ times more recovered stars, was much faster to run, and yielded consistent results (velocities within a median of $\sim2$ km/s of each other) with \ppxf\ for stars that were recovered in both \ppxf\ and \marz.

\subsection{Phase Space}
\begin{figure*}
    \centering
    \includegraphics[width=0.9\linewidth]{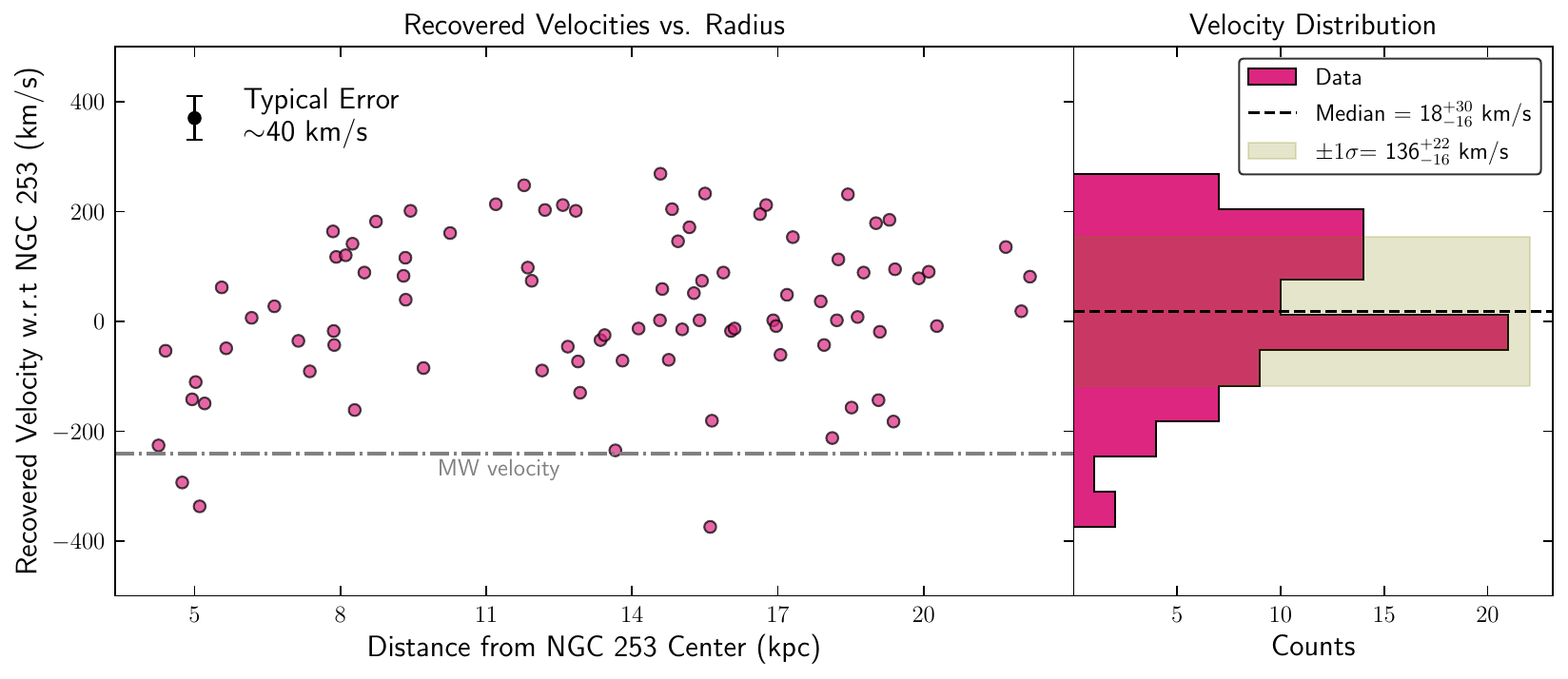}
    \caption{\textit{Left:} Distribution of stellar LOS velocities with respect to NGC 253's systemic velocity of 241 km/s of our recovered stars as a function of each star's projected radial distance from the center of NGC 253. The error bar shows the typical error of 40 km/s as derived from our mock spectra analysis. \textit{Right:} A histogram of the stellar velocities. The dashed line shows the median of the distribution, and the velocity dispersion, as calculated by the $1\sigma$ \texttt{mad\_std}, is shown as the shaded region.}
    \label{fig:phasespace}
\end{figure*}

We explore the phase-space distribution for each star in our final sample in Fig. \ref{fig:phasespace}, showing their LOS velocities relative to NGC 253's systemic velocity as a function of their distance from the center of NGC 253. A histogram of these velocities is shown in the right panel. The majority of our stellar velocities are much higher than the average MW foreground star LOS velocity (estimated as 0 km/s, so -241 km/s with respect to NGC 253's systemic velocity), indicating we do not have significant contamination from foreground MW halo stars. \urgent{See Sec. \ref{subsubsec:MWforeground} for further discussion of the effects of MW foreground contamination on our results.} We find the median velocity of all the stars in our halo sample to be $18_{-16}^{+33}$ km/s \katya{with respect to NGC 253's systemic velocity} and to have a velocity dispersion of $136_{-16}^{+20}$ km/s, with errors derived from bootstrapping. Median stacking all of our recovered spectra (without shifting back to the rest frame) and fitting the resulting stack with \ppxf\, we recover a LOS velocity with respect to NGC 253's systemic velocity of $39_{-21}^{+27}$ km/s and a velocity dispersion of $137_{-95}^{+48}$ km/s, which matches our calculations from individual stars within error.

\section{Seeking Velocity Substructure}\label{sec:discussion}

As we are interested in seeing if we can recover spatially coherent kinematic structure, we show the spatial distribution of our recovered velocities in Fig. \ref{fig:map}. Each point is color coded by the LOS velocity with respect to NGC 253's systemic velocity. Stars that were deemed unrecovered are marked as black x's. Though there are substantial outliers, we see hints that different regions of the halo may tend toward different characteristic velocities --- stars near the major axis tend to exhibit more positive velocities than those near the minor-axis, and there appears to be a cluster of stars with velocities hovering around 0 km/s in the SW. In order to check whether we can detect any velocity substructure, we attempt to smooth the velocity field.

\subsection{Locally Weighted Regression Smoothing}
\label{subsec:loess}
As our velocities have considerable error and outliers, we apply locally weighted regression (LOESS) smoothing to our spatial map to see if we detect any broad-scale trends in velocity. We use the implementation by \cite{Cappellari2013b_loess}, which uses the algorithm from \cite{cleveland1988locally_loess2d}. This technique fits multiple local regressions in subsets of points (``windows") across the dataset, thereby smoothing the data without assuming a specific functional form. For each window, the algorithm weights each point by its distance from the center of the window and performs regression on each window, resulting in a smoothed velocity field. \urgent{We use the default LOESS parameters, which use a linear local polynomial approximation and consider half of the points in the local approximation ($\texttt{frac}=0.5)$ which are weighted by their distance from each point in the regression. Errors are also calculated by the scatter of neighboring data points (\textit{i.e.} we do not use the \texttt{sigz} parameter) since we assume an identical error on each star.}

\begin{figure*}
    \centering
    \includegraphics[width=0.99\linewidth]{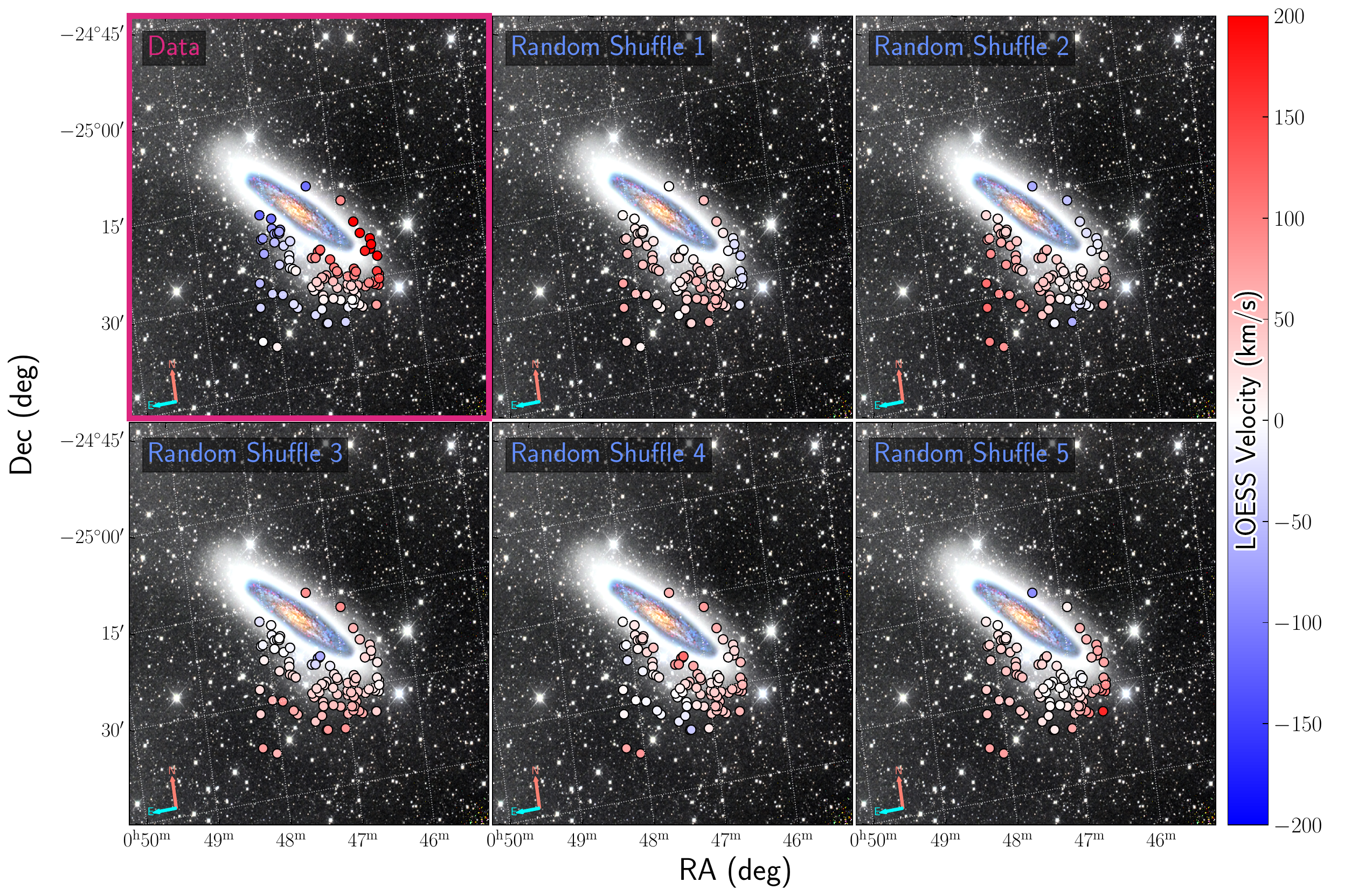}
    \includegraphics[width=0.99\linewidth]{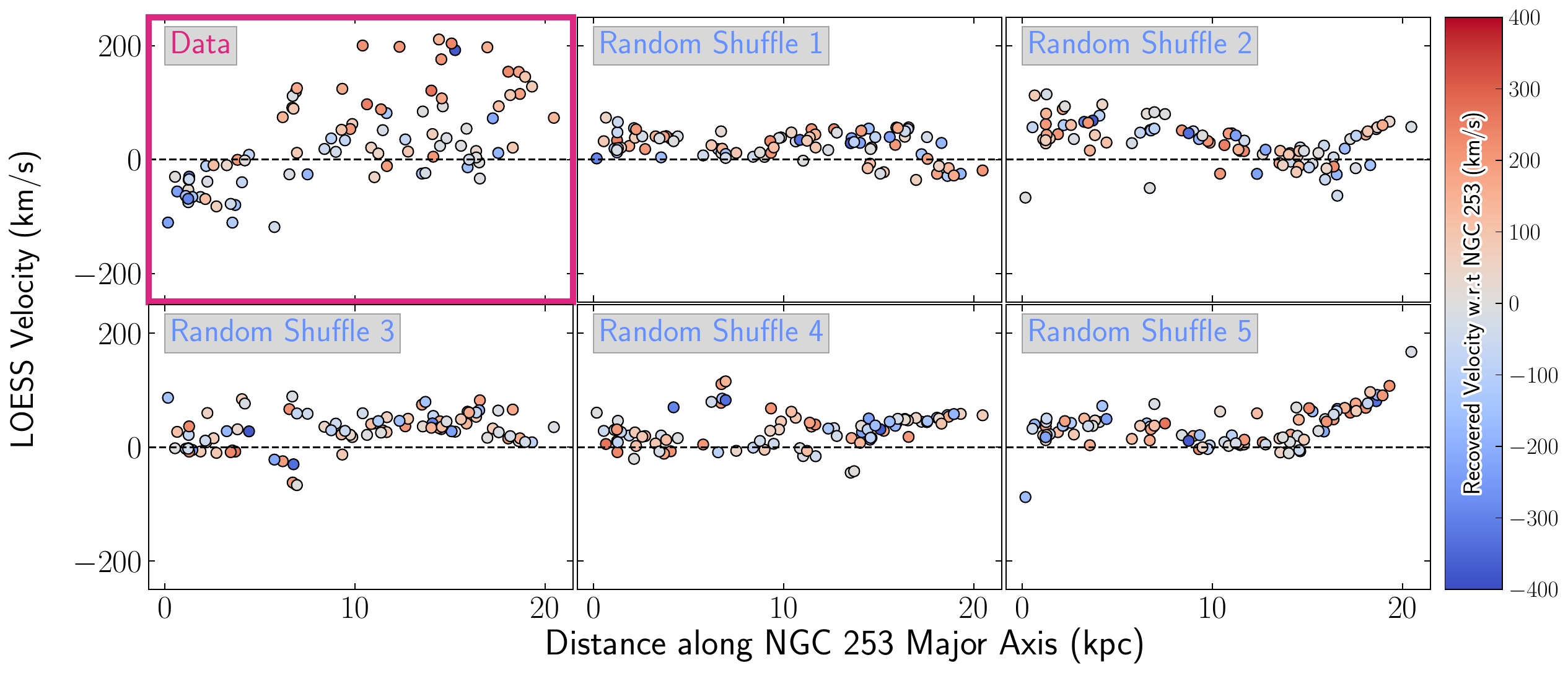}
    \caption{\textit{Top}: The same plot as Fig. \ref{fig:map} (top), but points are color coded by their LOESS-smoothed velocities. \textit{Bottom:} A similar plot to Fig. \ref{fig:phasespace}, except it is now showing the LOESS-smoothed velocities vs. the distance along the major axis rather than the radius. Points are color coded by their actual recovered velocities. In both top and bottom plots, the top-left panel shows the actual data LOESS smoothing, while the other panels show the result of LOESS smoothing 5 random shuffles of the velocities.}
    \label{fig:loess}
\end{figure*}

The outcome of this smoothing is shown in Fig. \ref{fig:loess}. The top-left panel of the top plot shows the LOESS-smoothed velocity field, while the top-left panel of the bottom plot shows a phase-space plot similar to Fig. \ref{fig:phasespace}, except this time it shows the LOESS-smoothed velocity against the \katya{projected} distance along NGC 253's major axis given a major axis position angle of 54$^\circ$, where points are color coded by their recovered velocities. The resulting spatial distribution in the top plot of Fig. \ref{fig:loess} shows a positive velocity along the major axis, negative along the minor-axis, and a cluster of stars with $\sim0$ km/s velocity to the SW, which we interpret as a sign of velocity substructure in the halo. \katya{We further investigate the nature of this substructure, including the classification of stars as \textit{in-situ} and accreted, in Section \ref{subsubsec:accvsinsitu}.}



\begin{figure}
    \includegraphics[width=1\linewidth]{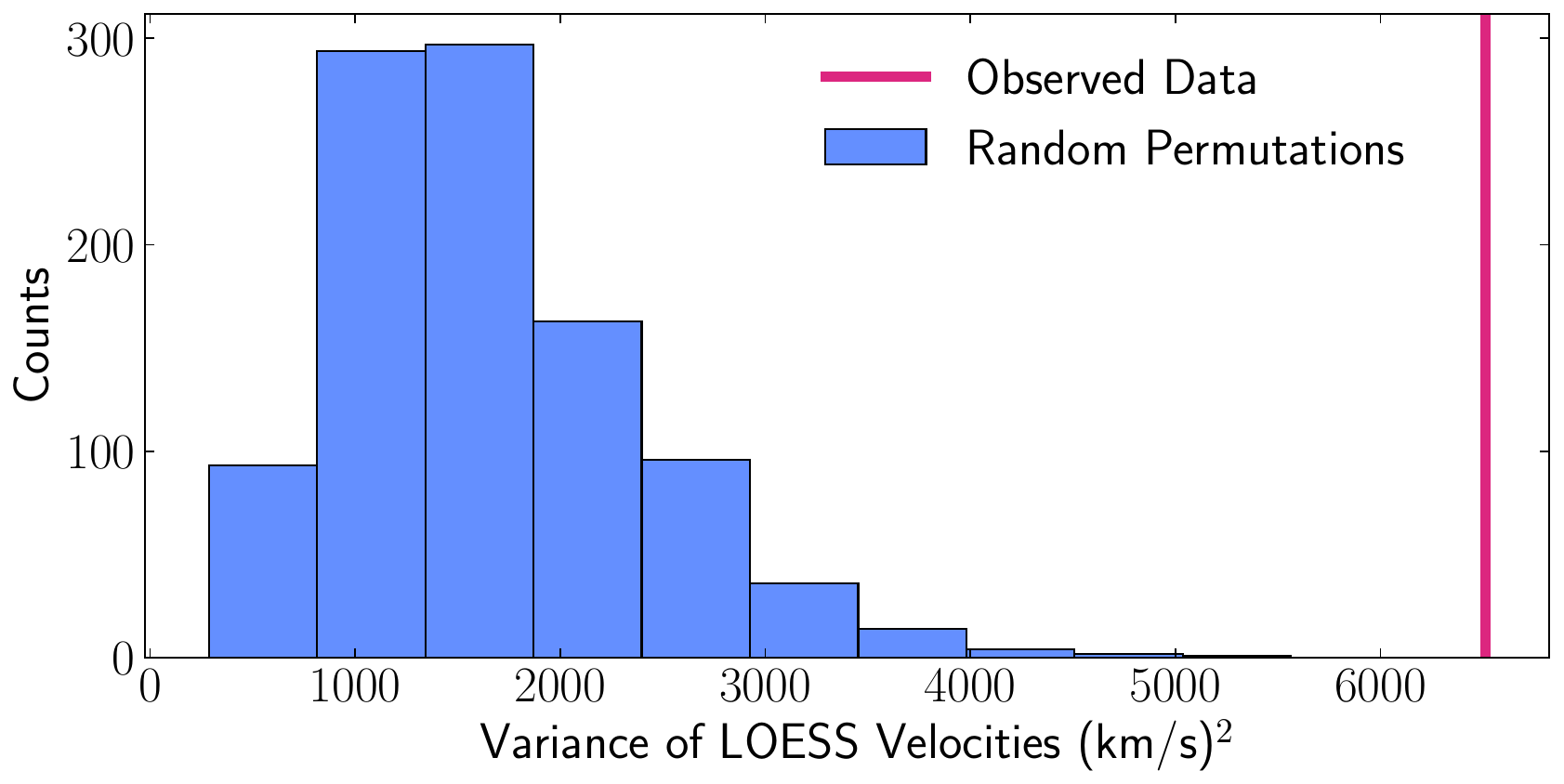}
    \caption{The variance of LOESS-smoothed velocities when randomly assigning velocities to stars at the positions of recovered stars (blue) and the variance of the actual data position-velocity distribution (pink). The actual data has a much higher variance in velocity than any of our randomly assigned distributions, indicating that we are capturing a signature of kinematic substructure that is unlikely to manifest in random (i.e. phase-mixed) distributions of velocities.}
    \label{fig:95perc}
\end{figure}

To check if the detected velocity gradient is genuine and is not the product of random scatter, we perform a similar experiment to the one we did with our stacked rest frame spectra. In order to keep the observed velocity distribution intact but break any relationship between velocity and position, we shuffle the velocities of all 89 stars and randomly reassign them to each star. We repeat this 1000 times and perform LOESS smoothing on these shuffled velocity distributions. The spatial maps and phase-space distributions of five random shuffles are shown in the remaining panels of Fig. \ref{fig:loess}. The random permutations of velocities show much smoother LOESS velocity fields than the data, and the phase-space plots generally show no correlation between radius and LOESS velocity for the random shuffles compared with the data, where one can see a broad positive velocity trend and considerable scatter. To compare with our actual spatial kinematics, we calculate the variance of the shuffled distributions and real velocities. A histogram and line showing the former and latter variances, respectively, are displayed in Fig. \ref{fig:95perc}. The real distribution has a higher variance than any of the 1000 random shuffles, indicating that it has more spatial coherence than expected by chance.

Therefore, we conclude that LOESS smoothing reveals statistically significant coherent kinematic substructure in NGC 253's stellar halo.

\subsection{M31: A LOESS substructure metric test case}
\begin{figure*}[!h]
    \centering
    \includegraphics[width=0.90\linewidth]{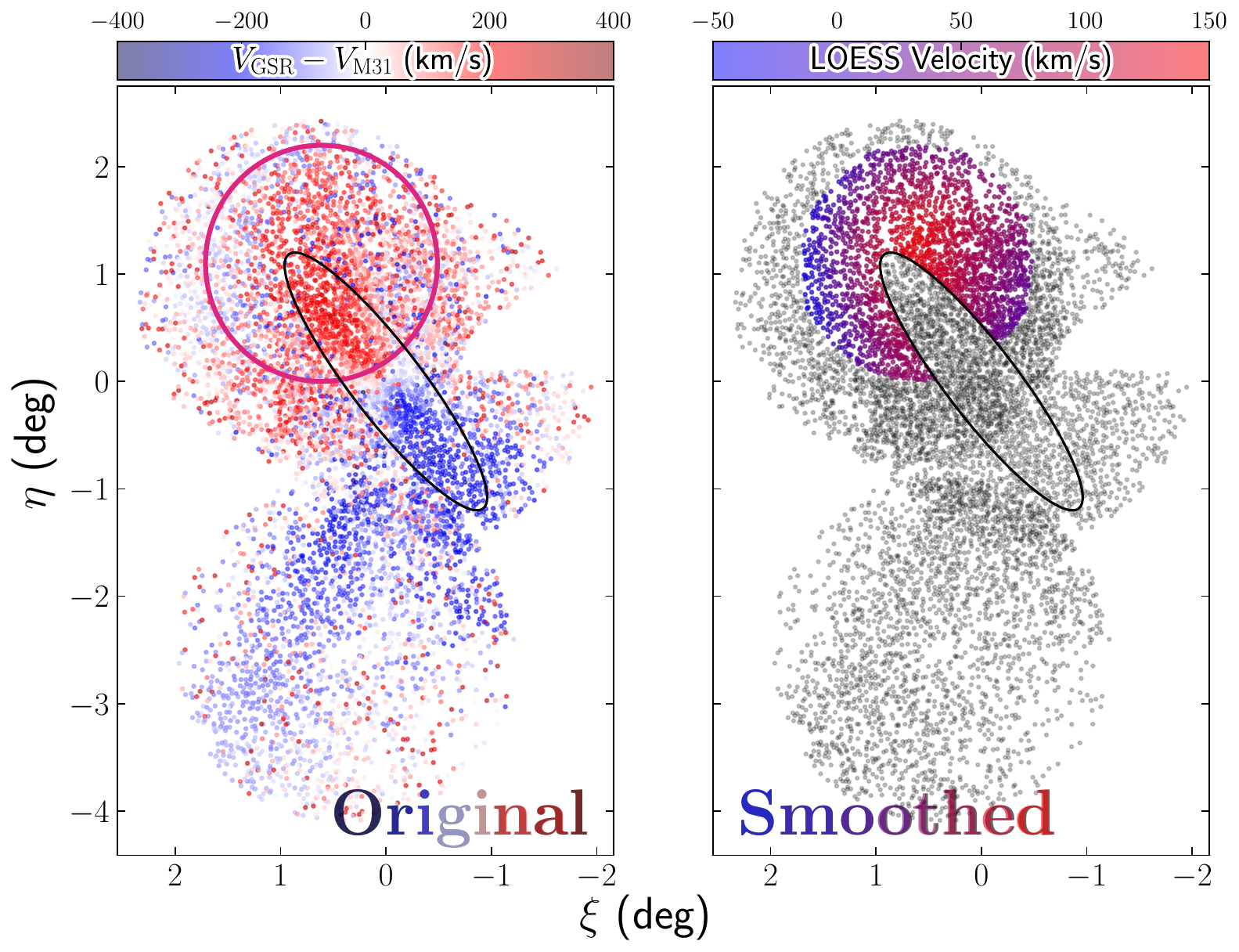}
        \includegraphics[width=0.90\linewidth]{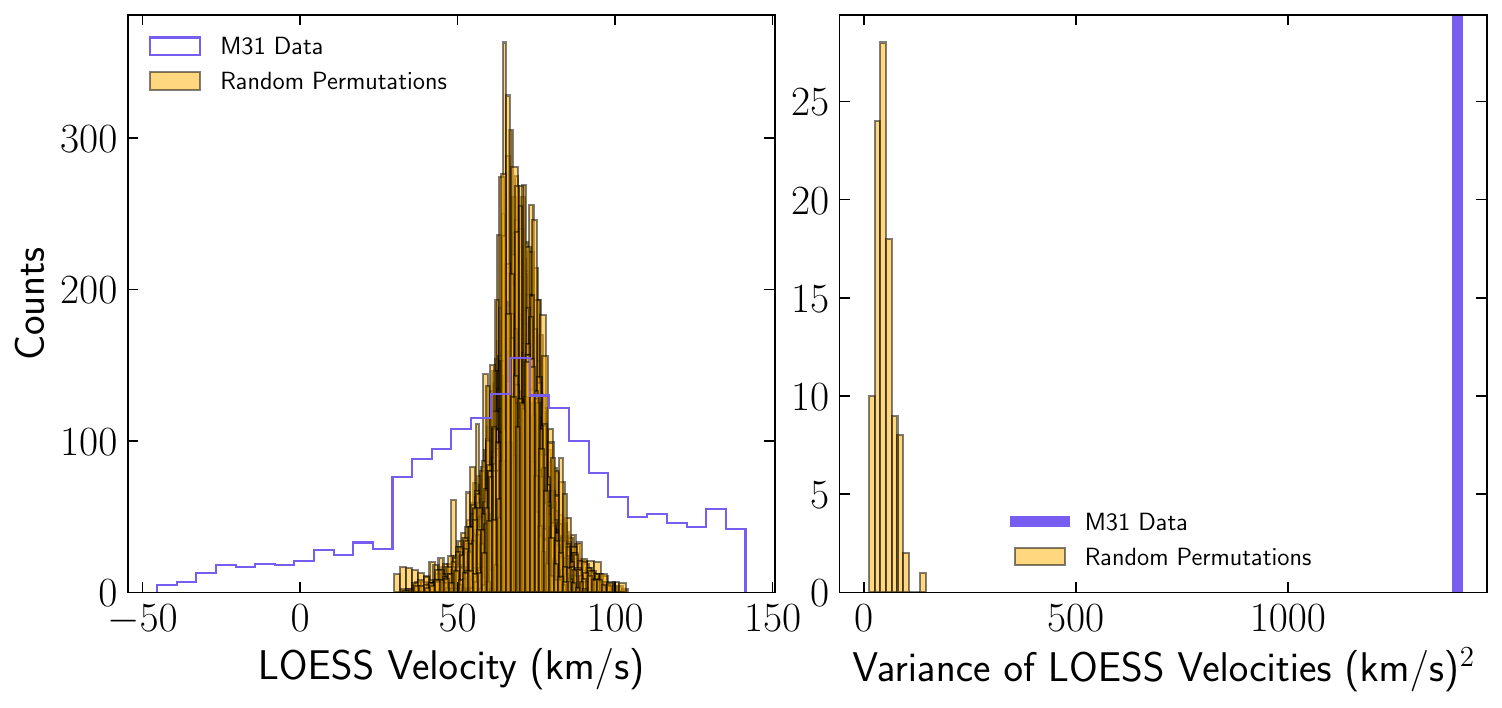}
    \caption{\textit{top-left:} A spatial map of stars in M31's halo, color coded by their LOS velocities from \cite{dey2023_desi} that have been converted to the Galactic standard of rest (GSR). A black oval is drawn around M31's disk, and a magenta circle around our region of interest that contains considerable substructure. \textit{top-right:} A spatial map where the stars in the substructure region are color coded by their LOESS-smoothed velocities. Note that the LOESS color bar scale has a much smaller dynamic range and different colormap than the one on the left. \textit{bottom-left:} The distribution of LOESS-smoothed velocities for the actual M31 data (purple) and random shuffles of velocities (yellow). \textit{bottom-right:} The variance of LOESS-smoothed velocities shown in the bottom-left panel. The M31 LOESS velocities (purple) have a much higher variance than any of the random shuffles (yellow), indicating the presence of spatially coherent structure.}
    \label{fig:m31}
\end{figure*}

We also repeated the same experiment on M31 in order to see if LOESS smoothing could indicate the presence of substructure. Andromeda has both a highly structured stellar halo and exquisite spectroscopic data of its halo stars from the DESI survey, making it the only observational analog to our method.  We used the LOS velocity data from \cite{dey2023_desi}. Since our NGC 253 observations focus on a specific region rather than the entirety of its halo, we also chose the NW region of M31's halo that mimics the M2FS FOV and contains part of its ``Northern Spur",  ``Northeast Shelf", and ``NGC 205 Loop" \citep{ferguson2016ASSL_m31substructurenames}. This region is highlighted in magenta in the top-left panel of Fig. \ref{fig:m31}. The LOESS smoothing of all the stars in the selected region is shown in the top-right panel of Fig. \ref{fig:m31}. One can see a clear velocity gradient that emphasizes the Northern Spur region. We performed the same velocity shuffling experiment as we did for our NGC 253 data, assigning random permutations of the available velocities to different stars in the region and LOESS smoothing the spatial map. The distribution of LOESS-smoothed velocities for the data and randomly shuffled sets is shown in the bottom-left panel of Fig. \ref{fig:m31}, and the variances of the distributions are shown in the bottom-right panel. This shows a very clear signature -- the M31 LOESS-smoothed velocities exhibit an extremely high variance compared to any of the random permutations, reflecting the structured and accreted nature of Andromeda's massive halo.

To estimate the impact of subsampling on the resulting LOESS velocities, we draw 51 random subsamples of 89 stars in our chosen region to mirror our NGC 253 selection and calculate the median difference between the LOESS velocities of the full sample and subsamples and the median error, which is given by the median of the 84th-50th and 50-16th percentiles. We measure a median offset of 2 km/s and a median error of $\pm30$ km/s. We also performed the shuffle test, performing LOESS smoothing on each of the 51 random subsamples and for 50 different velocity shuffles of each subsample. We find that the LOESS-smoothed actual velocities had a larger variance than the maximum variance of each shuffle 25\% of the time. Though subsampling does not significantly affect the calculated LOESS velocities of each star, subsampling can produce false negatives and mask the signal from existing kinematic substructure. Given M31's complex halo and accretion history, we acknowledge that there may also be structure on different scales that we do not capture with the size of the kernel we use for the LOESS smoothing.

\subsection{Spectral Clustering}
\label{subsec:spectralcluster}
As a further test of spatial structure, we perform an unsupervised clustering analysis on the three-dimensional phase-space coordinates (RA, Dec, $v$) of our recovered stars using the \textsc{scikit-learn} Python package \citep{pedregose2011JMLR_scikitlearn}. After comparing the behavior of different algorithms, we adopt spectral clustering (SC) via the \texttt{sklearn.cluster.SpectralClustering} function. SC is very well suited for our use case because it is nonparametric and makes no assumptions about cluster geometry, handling nonlinear and nonconvex structures well. Rather than relying on simple geometric distance, it identifies clusters by examining the connectivity structure of the data. 

Following \cite{vonLuxburg2007arXiv_spectralclusteringtutorial}, SC encodes our phase-space data as a similarity graph $A$ where each (RA, Dec, $v$) point is a node and the edges between nodes are weighted with a radial basis function (RBF), or Gaussian, kernel. The RBF kernel measures the similarity between points as a function of the Euclidean distance and a scaling parameter $\gamma$ factor that sets the kernel width: $A_{ij} = \mathrm{exp}(-\gamma ||x_i - x_j||^2)$. 
The algorithm then computes the Laplacian matrix of this graph as $L= D - A$, where $A$ is the similarity matrix as defined above, and $D$ is the degree matrix --- a diagonal matrix that for each star $i$ sums up the values of $\sum_j A_{ij}$, summing the similarity between star $i$ and all other stars $j$. The eigenvectors corresponding to the $k$ smallest eigenvalues of $L$ are extracted to project the data into a lower-dimensional, linearly separable space. The number of eigenvectors $k$ also sets the number of clusters and is a free parameter, which we select using the eigengap heuristic: the value of $k$ at which the difference between successive eigenvalues is maximized. We find $k = 3$ to be optimal, consistent with our by-eye inspection of the velocity field. Finally, \urgent{discretization (as implemented in} \textsc{scikit-learn}) is applied in this embedded space to assign the final cluster labels.

\begin{figure*}
    \centering
    \includegraphics[width=0.99\linewidth]{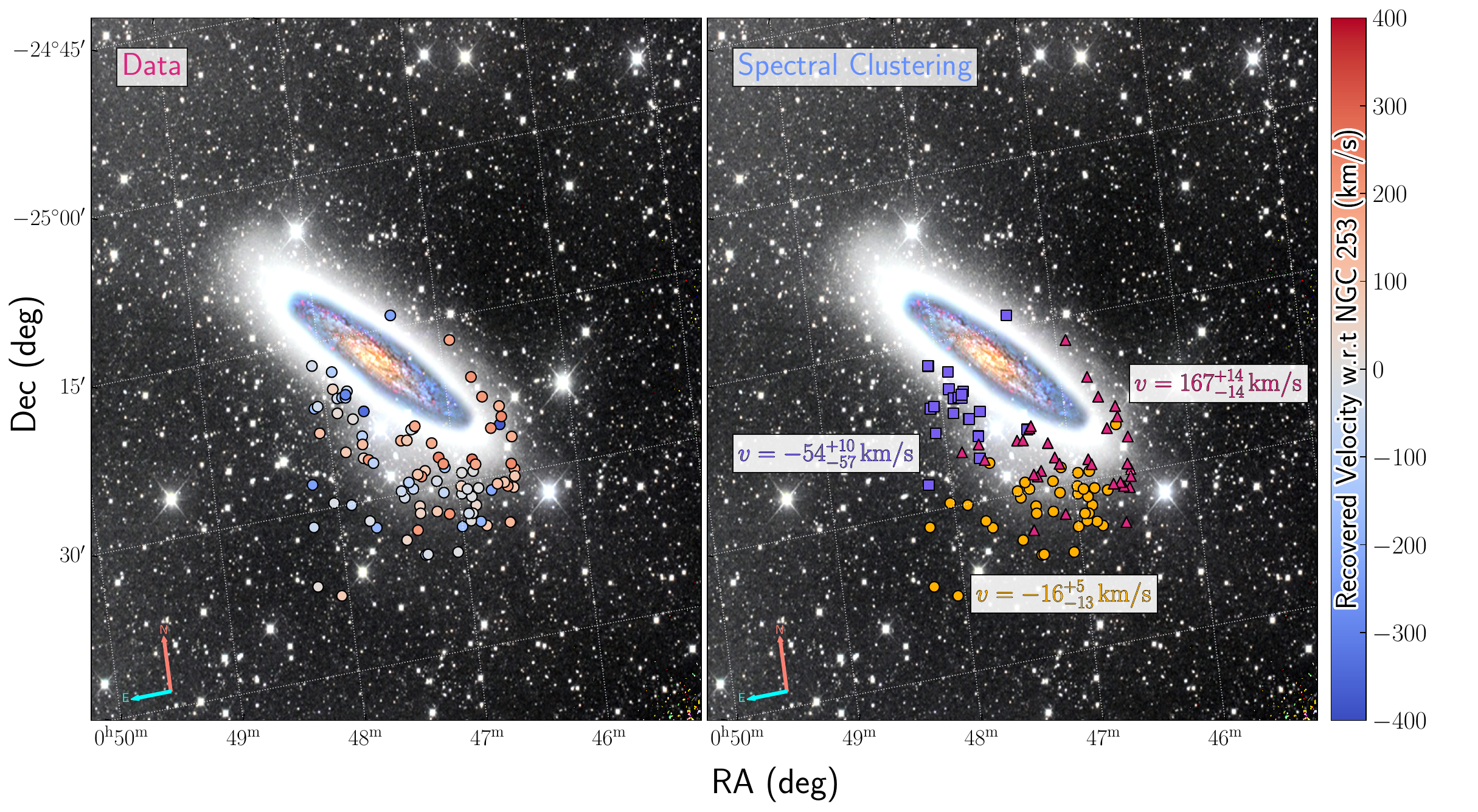}
    \includegraphics[width=0.99\linewidth]{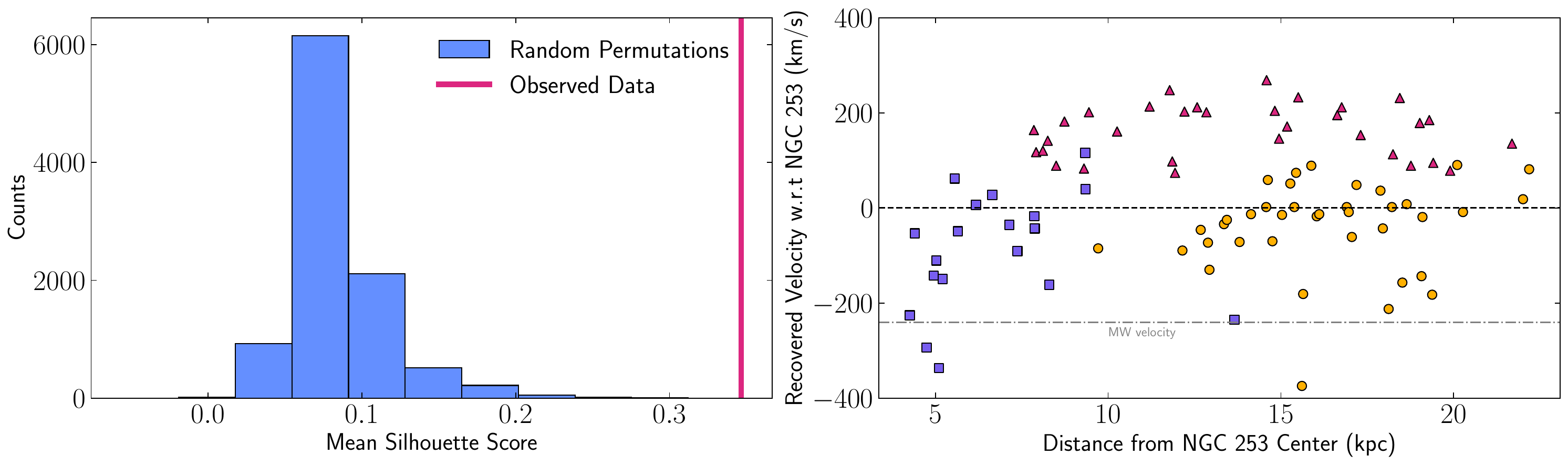}
    \caption{\textit{top-left:} Same as Fig. \ref{fig:map}, reproduced here for ease of reference. \textit{top-right:} Each recovered star is colored by the cluster it belongs to, as assigned by a Spectral Clustering algorithm. The median velocities of the clusters are displayed next to each cluster in the corresponding color. \katya{Note that all velocities are with respect to NGC 253's systemic velocity.} \textit{bottom-left:} The distribution of mean silhouette scores for the clustering of observed data (pink) and random permutations of velocities (blue). The observed data clustering has a higher silhouette score than any of the randomly shuffled trials, which indicates that the clustering found in the data is stronger than chance given our sampling. \textit{bottom-right:} The same phase-space plot as in Fig. \ref{fig:phasespace}, but simply color coded by cluster membership.}
    \label{fig:spectral}
\end{figure*}

Since our data have different units and scales, we first use \texttt{sklearn.preprocessing.RobustScaler} to scale our array of [RA, Dec, $v$], and then apply SC with three clusters. The identified clusters are visualized in the top-right panel of Fig. \ref{fig:spectral}. We also report the median velocity of the stars in each cluster \katya{with respect to NGC 253's systemic velocity} and errors by bootstrapping, and show the same phase-space plot of recovered velocity vs. radius as in Fig. \ref{fig:phasespace}, but color coded by cluster membership. We also perform a cluster significance test using the same technique as for LOESS smoothing. We perform 10,000 random shuffles of the velocities and run SC on these altered phase spaces. To compare to our data, we calculate the mean Silhouette score of each SC run for both the actual velocity field and the random shuffles. The Silhouette score subtracts the average distance between points in a cluster $d_{\mathrm{in}}$ from the average distance to points in the nearest neighboring cluster $d_{\mathrm{next}}$, divided by the maximum of $(d_{\mathrm{in}},d_{\mathrm{next}})$. 
The score ranges from $-$1 to 1, with a higher value indicating denser and better separated clusters. The mean Silhouette score of the observed data is higher than any of our 10000 random shuffles, indicating that there is a $<10^{-4}$ probability of our data being drawn from an unclustered dataset.

Examining the identified clusters, we see that they align well with the structure found by LOESS smoothing. While the cluster visualized as purple squares in Fig. \ref{fig:spectral} lies along the minor-axis of the galaxy and has a fairly negative median velocity of $-54_{-57}^{+10}$ km/s, the magenta triangle cluster that lies along the major axis has a large positive median velocity of $167_{-16}^{+14}$ km/s. The remaining yellow circle cluster in the SW corner has a median velocity of $-16_{-13}^{+5}$ km/s.

\begin{figure*}
    \centering
    \includegraphics[width=0.99\linewidth]{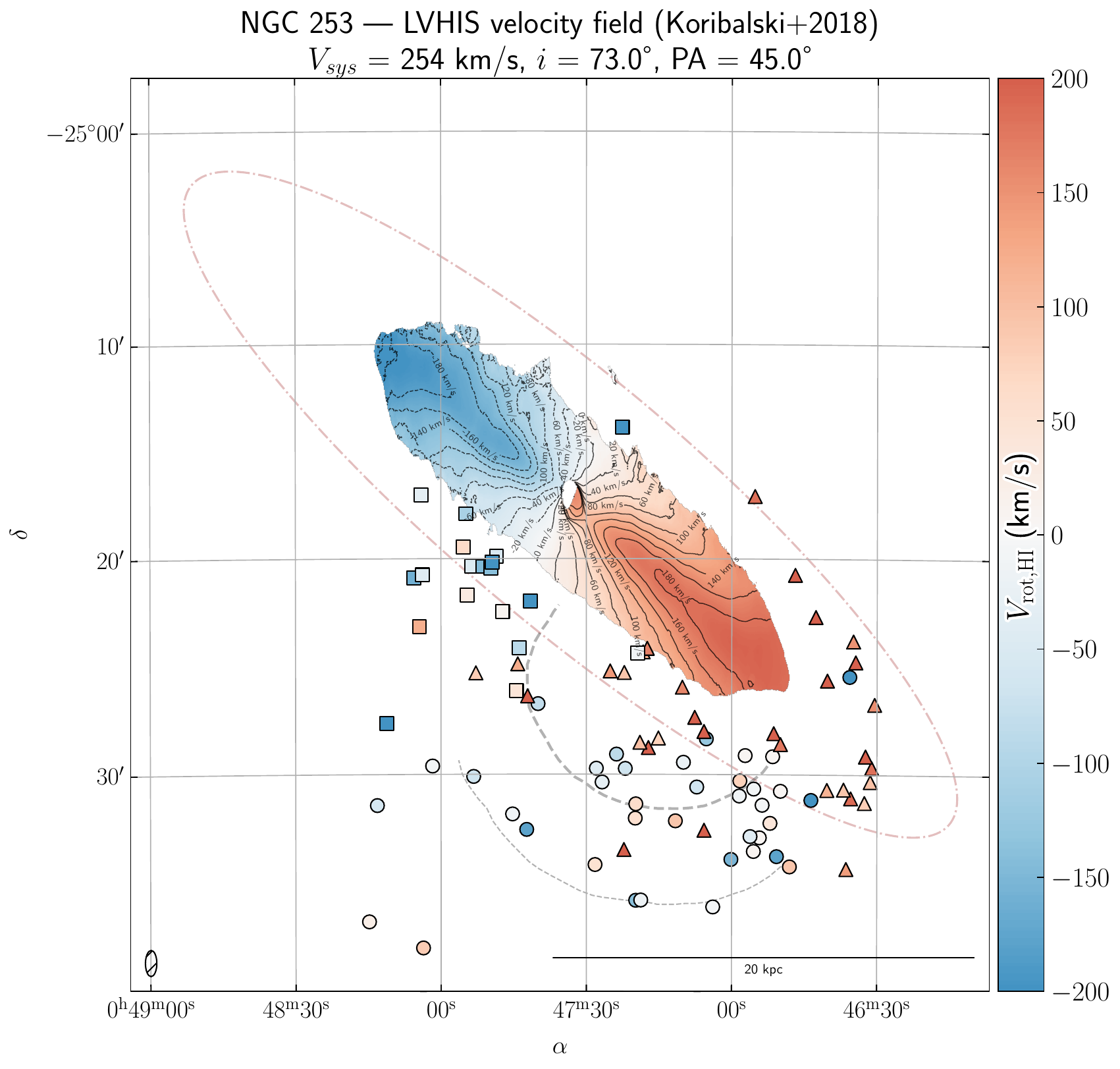}
    \caption{\katya{The HI rotation velocity map from \cite{koribalski2018MNRAS_LVHIS}. The beam size is show in the bottom-left corner as a hatched oval.} 
    The stars recovered in this paper are shown as dots, color coded by their recovered velocity in the same color scaling as the gas velocity. The point markers correspond to the cluster they are assigned by the Spectral Clustering algorithm as discussed in Sec \ref{subsec:spectralcluster}. All our measured velocities are relative to NGC 253's systemic velocity. \urgent{Dashed gray contours show the approximate location of the shell (same as Fig. \ref{fig:map}). The brown dashed-dotted line is an ellipse with a major axis radius of 25 kpc and minor-axis radius of 6 kpc which represents an approximation of the outer disk from \citealt{Greggio2014}}}.
    \label{fig:HIfield}
\end{figure*}

\subsubsection{Accreted and \textit{in-situ} components}
\label{subsubsec:accvsinsitu}
Our SC results are in strong agreement with the velocity substructure identified by LOESS smoothing, with both methods recovering the same kinematic components despite a significant velocity error of $\pm40$ km/s and a modest number of stars. There exists a population of stars along the minor-axis with negative velocities, a population along the major axis with large positive velocities, and a kinematically cold group near zero velocity concentrated in the SW, spatially coincident with the known shell feature. For both methods, permutation tests show that the observed level of spatial coherence is extremely unlikely to arise from randomly assigned velocities. The agreement between two distinct methodologies -- spatial smoothing and unsupervised graph-based clustering -- lends confidence to the conclusion that we have detected clear kinematic substructure in NGC 253. 

\katya{However, SC does not tell us the nature of each cluster, only that it has detected them. \urgent{In this subsection}, we attempt to create a physically motivated interpretation and distinguish between accreted and \textit{in-situ} components, which can both be present in a stellar halo.}

\katya{In fact, we believe that we have detected both in NGC 253. 
The smoothed velocity along the minor-axis of the galaxy is significantly negative, while one would expect these velocities to be close to zero if these stars were simply rotating \textit{in-situ} disk stars. Additionally, the region that has near-zero rotation is offset from either the minor or major axis to the southwest part of the halo near where the shell is. To understand how common it is to see nonzero velocities on the minor-axis, we look at simulated halos from the TNG-50 simulation to understand the spatial distribution of stellar velocities in the central regions of MW-mass halos. We find that only 0.8\% of the 471 halos analyzed have accreted star $|v|>50$km/s in the central 7.5 kpc, and that all of the halos have \textit{in-situ} star velocities $\sim0$ km/s, making our NGC 253 measurement an outlier in this regard. 
The distinct spatial distribution of velocities and the rarity of comparable minor-axis central velocities in simulated halos support the interpretation that part of the stellar halo bears a clear kinematic imprint of accretion.}

\katya{While the pronounced nonzero velocity cluster along the minor-axis and the localized, offset low-velocity feature are inconsistent with an \textit{in-situ} disk and strongly point to being a substantial accreted contribution, the same cannot be immediately argued for the major axis cluster. Indeed, upon examination of the median cluster velocities, we notice that the stars in the major axis cluster present generally high velocities that could be consistent with a disk velocity.}

\katya{As an initial simple experiment, we see if the major axis cluster stars could be drawn from a different stellar population than the rest of our stellar sample by performing a two-sample Kolmogorov-Smirnov (K-S) test on the colors (as a crude proxy for metallicity) of the major axis cluster stars vs. the remaining stars with measured velocities. The resulting p-value is 0.52, meaning that we fail to reject the null hypothesis and find that they are not drawn from different underlying stellar populations.
}

\katya{As a more rigorous test of whether the different clusters we have detected are more likely accreted or \textit{in-situ} in origin, we compare with the disk rotation velocity field. We obtained a map of the HI gas velocity field from \cite{koribalski2018MNRAS_LVHIS}, as well as data tables from three different sources that measured the gas velocity rotation curve of NGC 253: \cite{puche1991AJ_HI} and \cite{lucero2015MNRAS_HI} (using HI) and \cite{HlavacekLarrondo2011} (using H$\alpha$). We note that, unusually, NGC 253's HI gas disk is truncated and does not extend as far as its stellar disk, which has been attributed to possibilities such as starburst ionization or stripping \citep{boomsma2005A&A_NGC253HIextraplanar}.}


\katya{We illustrate the HI velocity field of NGC 253 in Fig. \ref{fig:HIfield} using the velocity (moment 1) map from \citealt{koribalski2018MNRAS_LVHIS}. We also plot the recovered velocities of our stars using an identical color scheme, and we mark each measurement by which cluster it belongs to from the SC algorithm. It is immediately evident that the stars in the major axis cluster (triangles) have radial velocities that are mostly consistent with the gas disk rotation velocities, unlike the other two identified clusters.}

\begin{figure*}
    \centering
    \includegraphics[width=0.99\linewidth]{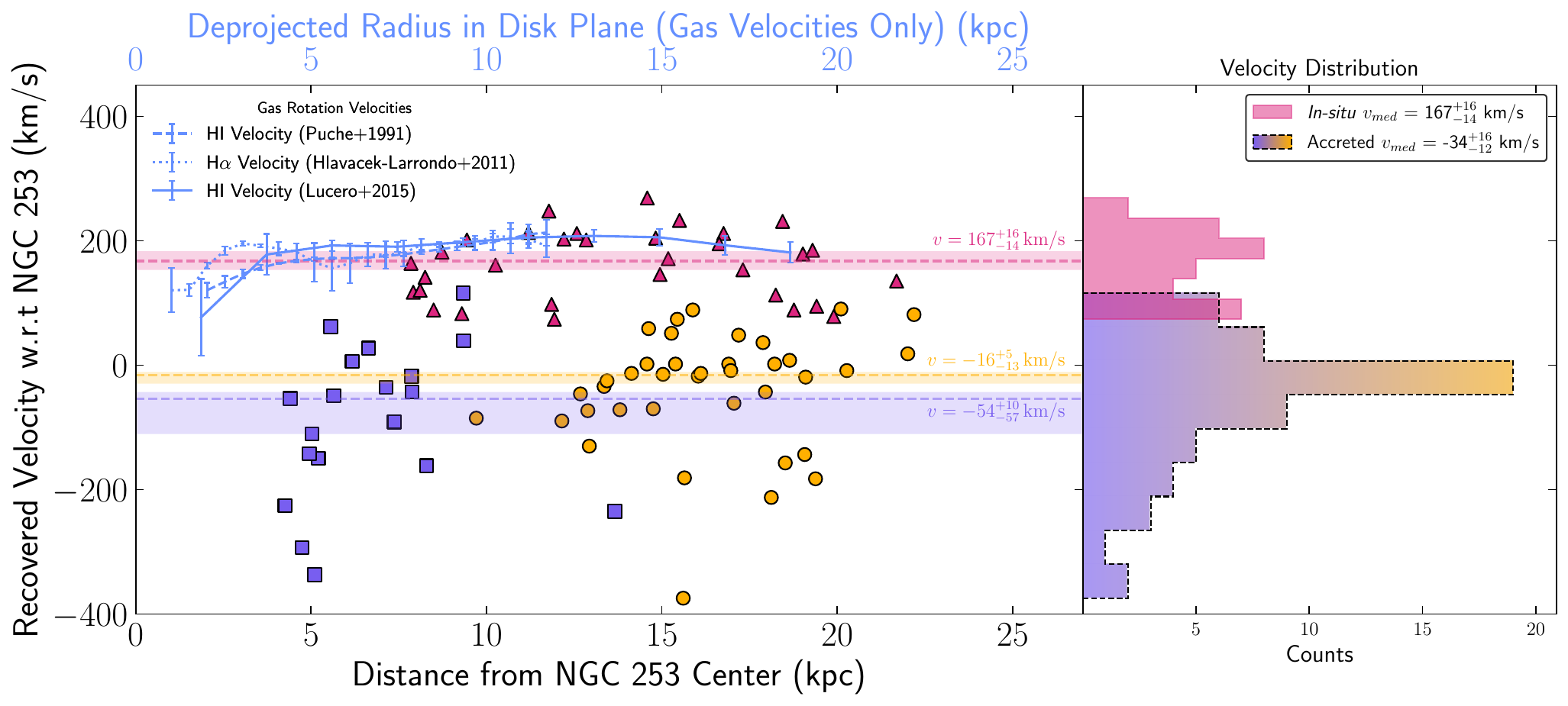}
    \caption{\katya{\textit{Left:} The same phase-space plot as in Fig. \ref{fig:phasespace}, but simply color coded by cluster membership. Dashed horizontal lines and shaded regions show the median velocity of each component and its $\pm1\sigma$ uncertainty. Blue lines of various styles show the neutral gas rotation curves from \cite{puche1991AJ_HI, HlavacekLarrondo2011} and \cite{lucero2015MNRAS_HI}. Note that the velocities and radii of the rotation curves correspond to a deprojected radius in the disk plane, which differs from the radius from the center of NGC 253 that is being shown for the stars because the disk is highly inclined. Nevertheless, we see the striking similarity between the gas rotation velocities and the radial velocities of the major axis stars (magenta triangles). \textit{Right:} Histograms of the velocities of stars we identify as being either \textit{in-situ} or accreted components. 
}}
    \label{fig:gasrotation}
\end{figure*}

\katya{The left side of Fig. \ref{fig:gasrotation} shows these neutral gas rotation curves overplotted on the same phase-space diagram as at the bottom-right of Fig. \ref{fig:spectral}. Note that due to NGC 253's high inclination angle of $76^\circ$, the velocities and radii cited for the rotation curves (top x-axis in blue) correspond to a deprojected radius in the disk plane, which is distinct from the radius we plot each of our stellar velocities at (bottom x-axis in black). Nonetheless, even this rough comparison shows that the majority of stars in our major axis cluster appear to have velocities that are consistent with the gas velocity, making it likely that these stars actually trace the \textit{in-situ} outer disk of NGC 253 to quite large radii rather than an accreted halo.}

\katya{Assuming that the minor-axis and offset regions are likely accreted while the major axis region is likely \textit{in-situ}, we calculate the velocities \katya{and standard deviations (both simple standard deviation (STD) and \texttt{mad\_std})} of these two components. We show histograms of stellar velocities in these regions on the right side of Fig. \ref{fig:gasrotation} and find that the median line-of-sight velocity of the \textit{in-situ} population is $v_{med, in-situ} = 167_{-14}^{+16}$ km/s with a $\sigma_{\rm \textit{in-situ}, STD} = 53 \pm 5$ km/s and $\sigma_{\rm {\textit{in-situ}, MAD\_{STD}}} = 60 \pm 13$ km/s. This velocity is quite high compared to the velocity of the accreted population, which is modestly negative with $v_{med, acc.} = -34_{-12}^{+16}$ km/s, $\sigma_{\rm accreted, STD} = 104 \pm 12$ km/s and $\sigma_{\rm {accreted, MAD\_{STD}}} = 82^{+24}_{-16}$ km/s. In fact, this high velocity would be quite an outlier if the major axis stars were accreted; $<0.5\%$ of 471 MW-mass galaxies in TNG-50 have accreted velocities at r=15 kpc that are $\geq167$ km/s, indicating that accreted velocities of the magnitude seen along the major axis of NGC 253 are uncommon.}



\katya{Given this evidence from neutral gas rotation velocities and simulations, we believe that the cluster of major axis stars we detected are likely mostly \textit{in-situ} disk stars, while the minor-axis and off-axis regions are mostly accreted halo and an accreted shell, respectively. Though we acknowledge that due to our small sample size, the velocities of the accreted halo and accreted shell components are similar and may be drawn from the same distribution, we choose to interpret them as two distinct components for a few reasons. Despite the shell and halo having similar median velocities, they still differ by $>1\sigma$ from each other, and the distribution of accreted shell velocities produces a noticeable peak in the marginal histogram. More importantly, the eigengap heuristic of SC detects three components, including a distinct component that is remarkably spatially coincident with the known photometric shell. This component was recovered independently of any knowledge of photometric substructure and was purely based on a combination of positions and velocities, which shows evidence for clustering that is not apparent in velocity distributions alone. Nonetheless, given the \urgent{small} number of stars and modest velocity accuracy, we view this detection of substructure as somewhat tentative. For the very limited number of halos that have spectroscopically resolved stellar measurements, it is clear that additional data (e.g. larger samples, chemistry) are quite helpful in enabling the decomposition of halo stars into clearer and more detailed substructure \citep{naidu2021ApJ_MWmajormergerH3survey, escala2023AJ_SPLASH_M31Inner,dey2023_desi}. Accordingly, we anticipate that future wide-field instrumentation will further clarify the nature and extent of kinematic substructures in the NGC 253 halo.}

\katya{Since the majority of our major axis stars are likely \textit{in-situ}, we cannot make any definitive statements about the overall halo velocity or its AM (e.g. whether it is prograde or retrograde). That said, we are uncertain about the nature of any particular star --- a subset of our \textit{in-situ} stars could be accreted and phase-mixed, which would imply that the halo could have prograde rotation. We briefly speculate on the implications of a prograde rotating halo in Sec. \ref{subsec:AM}; future higher-resolution spectroscopy that could measure metallicities or abundances can help distinguish between different stellar origins.}

\subsubsection{Possible MW foreground contamination}
\label{subsubsec:MWforeground}
\urgent{As noted previously, there is a possibility that some of our highly negative velocity stars may be consistent with being MW foreground contamination. There are 12 stars in our sample with velocities within 100 km/s of the MW velocity (-241 km/s, in the frame of NGC 253's systemic velocity). In order to test how these stars might impact our interpretation of substructure, we perform two experiments.}

\urgent{In the first, we see whether the spatial distribution of these 12 stars is consistent with MW foreground, assuming that such stars should be uniformly scattered around our FOV. We randomly distributed 12 points around our FOV and counted the number of points that fell within a 30' $\times$ 5' box aligned with the minor-axis of NGC 253 (assuming a major axis PA = 53$^\circ$). In 10,000 trials, the mean percentage of stars in the minor-axis region was $\sim$14\% ($\sim$1.66 stars), whereas 6/12 (50\%) of our stellar sample are in this minor-axis region. The chance of having six or more uniformly distributed sources fall in the minor-axis is 0.28\%, indicating that there is a statistically significant difference between the number of stars that appear in the minor-axis region when uniformly distributed and the number of possible MW foreground stars in this region that we see in our data. This makes it very unlikely that \textit{all} of our highly negative velocity stars are MW foreground interlopers --- some could be \textit{bona fide} NGC 253 stars with MW-like velocities.}

\urgent{Given this, we performed a second test to see what excluding stars that may be MW interlopers would do to our clustering analysis. Out of the 12 possible MW foreground stars, we pessimistically exclude eight stars --- all six sources outside of the minor-axis, and two stars at random on the minor-axis (as per our previous experiment) --- and run clustering again. We find that SC still prefers three components even while excluding these stars, and only 1-2 of the remaining stars, usually on the edge between labels, change their classifications. The resulting median velocities of each of the three clusters are also consistent (within error) with the originally identified cluster velocities. Therefore, we conclude that the presence of possible MW interlopers in our stellar selection does not significantly alter the conclusions of the paper.}


\section{NGC 253's Halo Kinematics}

Using M2FS spectroscopy of 89 faint halo stars in NGC 253, we have clearly detected \katya{stars} with a modest median velocity of $\sim18$ km/s and dispersion of $\sim136$ km/s, which matches within error with the \ppxf-measured velocity and velocity dispersion of a median stack of all recovered spectra. By smoothing and clustering the spatial distribution of these stars, we have detected kinematic substructure in the halo of NGC 253.  

We emphasize that this is the first-ever detection of its kind for a MW-mass galaxy outside of the LG and the fourth-ever measurement of an overall halo velocity for any MW-mass system. \katya{Through unsupervised clustering, we find three clusters of stars --- a negative velocity region along the minor-axis, an almost-zero velocity offset region coincident with NGC 253's photometric SW shell, and a highly positive velocity region along the major axis. By comparing with neutral gas velocities in NGC 253's disk and also simulated MW-mass galaxies from TNG-50, we infer that the region along the major axis with high LOS velocities is mostly part of an \textit{in-situ} outer disk, while the other minor-axis region is more consistent with being part of an accreted halo and the offset region likely tracing an accreted shell.}

\katya{\cite{iodice2014A&A_ngc253IRIVSTA} and \cite{Greggio2014} use VISTA to find that NGC 253 has an exponential profile out to quite large radii $(\sim$ 20 kpc) in the J-band. They attribute this to an extended outer disk, though there are examples of galaxies with halolike stellar populations that follow exponential profiles \citep[e.g. M101;][]{jang2020A&A_m101halo}. By measuring the kinematics of NGC 253's major axis stars for the first time, our work provides additional evidence that clarifies the nature of some of these stars --- they are indeed likely part of a rotationally supported outer stellar disk that extends out to considerably large radii.} 

\katya{This separation of stellar components underscores the importance of combining photometric detections of halo stars with stellar kinematic measurements. Photometry provides information about the stellar populations present in the halo, including color and photometric metallicity estimates, but much stellar halo work is confined to measuring only the minor axes of highly inclined halos where \textit{in-situ} contamination is assumed to be relatively negligible \citep[e.g.][]{Harmsen2017, smercina2020m81}. Rarely, the disk and halo populations are sufficiently distinguished in wide-field imaging covering an entire halo or in face-on systems such that one can isolate the halo contribution \citep[e.g.][]{gozman2023ApJ...947...21G}, but in many cases stellar-population information alone is not sufficient to definitively differentiate the two. Spectroscopy of these stars, though currently challenging, presents us with a more robust way to differentiate accreted versus \textit{in-situ} stars, especially in cases where it is not apparent from stellar populations alone.}

In the following sections, we aim to connect our measurements with NGC 253's merger history and put them into context with previous literature. We explore markers and implications of dynamical youth in its halo, \katya{followed by briefly theorizing the implications of being able to measure halo AM and exploring its relation to disk orientation. } 
We then zoom out and compare what we have learned about its dynamical youth and AM with the presence of those signatures in other nearby galaxies, and finish by discussing some caveats of our work and what the future may hold for halo kinematic analyses. 

\subsection{NGC 253's Dynamical Youth and Implications for Halo AM}
\subsubsection{Connection to the shell and implications for a recent accretion}
Many studies present structural signs that NGC 253 had a recent encounter with a companion. For one, there is evidence that NGC 253 possesses a lopsided and disturbed disk \citep[see also][]{Greggio2014, lyu2023MNRAS_NGC253diskasymmetry} with differing levels of star formation and UV emission, and that it hosts regions of ``extraplanar stars," \katya{which may be consistent with a tidal encounter from a companion close to the disk plane \citep{davidge2010ApJ_ngc253disrupted}.} Additionally, the substructure of NGC 253's halo that has zero velocity overlaps with the region where a shell has been detected in integrated light \citep{malinandhadley1997}, and out to larger radius with resolved stars \citep{davidge2010ApJ_ngc253disrupted, bailin2011, Greggio2014} where it appears as an overdensity in both spatial maps and stellar density profiles from star counts and has been attributed to the possible past merger of a relatively massive galaxy. 
Given that it is detected both photometrically and spectroscopically, and is mostly present in the inner halo where the dynamical phase mixing timescale is shorter, this indicates that it arose from a relatively recent ($\sim$few Gyr) accretion event. Ascertaining the exact timing of the accretion would require comparison with simulations and dynamical modeling and deeper work in understanding star formation histories of nearby galaxies, which is beyond the scope of this paper, but may be investigated in future work. Nonetheless, our work points to NGC 253 having an important merger fairly recently, which is also hinted at in other studies.


Stellar-population diagnostics provide a complementary (though model-dependent) window into halo assembly and youth. \cite{harmsen2023MNRAS_agbrgb} calibrate the ratio of AGB-to-RGB stars from HST/ACS GHOSTS \citep{radburn-smith2011ApJS_ghosts_ngc253dist} fields in nearby galaxies to correlate it with their star-formation histories through t$_{90}$, the time before which 90\% of the stars formed. They find a t$_{90}$ for the inner halo of NGC 253 to be $6.2\pm1.5$ Gyr and less ($\sim 5$ Gyr) for its overdense shelf at 30 kpc in the NW minor-axis, though with overlapping uncertainties and with a number of limitations and calibration uncertainties associated with this relatively new technique. Though they use two fields to define the ``inner halo" and consider it a proxy for the ``well-mixed" halo, their fields are quite small (3.4' $\times$ 3.4' \urgent{or $\sim 3.5 \times 3.5$ kpc assuming D=3.5 Mpc}), and one of their inner halo fields overlaps with part of the SW shell region that we targeted, but they decide to include it as they did not find variation in the AGB/RGB ratio between the two fields. In the future, the High-Latitude Wide-Area Survey of the Roman Space Telescope will enable the measurement of AGB/RGB ratios across the entire halo field of NGC 253 and other Local Volume galaxies, letting us place better constraints on the star formation and stellar populations of nearby halos \citep{sanderson2026arXiv_Romanearfield}.

GCs may also encode signatures of youth. \citet{cantiello2018A&A_NGC253GCs} looked at properties of GC candidates around NGC 253 with VST and VISTA and found that candidates near the shell region identified by \cite{bailin2011} are, on average, bluer than GCs in their confirmed sample and in other randomly drawn regions (though within 1$\sigma$ of the median colors of their confirmed sample due to small-number statistics) and hypothesize that these may be remnants from an accretion with a low-mass satellite galaxy. They also find an asymmetric globular cluster luminosity function (GCLF) that seems to be missing bright GCs. Beyond proposing that such systems may be hidden in the dusty disk of NGC 253 or located at large radii, they speculate that their dearth of bright GCs may also be linked to NGC 253's dynamical youth, as a recent ($\sim 6$ Gyr) accretion may have brought in some intermediate-age clusters that have not disrupted yet through tides and would be overrepresented in the GC population, hence skewing the GCLF to higher luminosities.



\subsubsection{How satellites influence disk AM}
\label{subsec:AM}

\katya{Simulations predict that stellar halos are built from a few of the most massive progenitors \citep{monachesi2019, Khoperskov2023A&A_Hestiahalo, monzon2026arXiv260118954M}; therefore, halo properties can reflect the properties of its dominant mergers and encode the effects of these mergers on the galaxies themselves.}

\katya{One of these effects is disk tilting. As mentioned before, simulations show that the majority of galaxy disks reorient due to satellite accretion \citep{gomez2017MNRAS_aurigadiskwarpreorient, dillamore2022MNRAS_ARTEMISdisktilt, bell2026OJAp_mergersAM}. This is especially prominent in younger accretions; \urgent{\cite{bell2026OJAp_mergersAM} find that the mean velocity of accreted halo stars at 35 kpc along the major axis is strongly positively correlated with both (1) the component of the stellar halo AM aligned with the \textit{in-situ} AM and (2) the component of the dominant merger specific AM aligned in the direction of the \textit{in-situ} star AM.} The latter relation is quite pronounced for mergers that happened more recently than $\leq 7.2$ Gyr ago compared to older mergers, where the correlation is almost imperceptible.}

\katya{Unfortunately, our present-day knowledge of NGC 253's halo  makes it difficult to connect it to the kinematics of its dominant merger. Its halo major axis is likely dominated by \textit{in-situ} stars, and we do not attempt to attribute any individual star as being accreted or \textit{in-situ} given the large uncertainty and lack of finer chemo-dynamical knowledge, so we do not make definitive statements regarding NGC 253's halo AM. Our analysis shows that its halo does appear to contain substructure we believe was accreted (e.g. its shell, which is part of the halo and off-center), and this substructure does not appear to be rotating in the same direction as its disk. Given NGC 253's apparent dynamical youth, this may seem a little unusual given that disk reorientation is especially prominent in younger accretions, so it may be an interesting example of a disk that did not experience reorientation due to merger. This may make NGC 253 a compelling test case for understanding disk tilting due to merger, and future wider-area spectroscopy of this system and other halos would form a particularly accessible observational record of this phenomenon.}  



\subsection{Comparison with Other Stellar Halo Kinematics}

There are only three other galaxies we know of with resolved-star halo spectroscopy -- the MW, M31, and NGC 4945. Although differences in sample size, FOV, and location in the halo complicate direct comparisons, we attempt to contextualize our measurements and inferences of dynamical youth and AM within this set.

\subsubsection{Signposts of Dynamical Youth}
There is a stark contrast between the dynamical maturities of the MW and M31, which have very different accretion histories. 

The 6D phase space of individual stars in the MW's highly anisotropic stellar halo have been richly explored by Gaia (\citealt{gaia2016A&A...595A...1G}) and large spectroscopic surveys such as the SDSS SEGUE and APOGEE surveys \citep{yanny2009AJ_SEGUE, majewski2017AJ_APOGEE}, the H3 survey \citep{conroy2019ApJ...883..107C_h3}, GALAH \citep{martell2017MNRAS_galah}, and DESI \citep{DESI2025arXiv250314745D}. This abundance of surveys found that its halo contains a $\sim0.06$ solar metallicity anisotropic component stemming from the debris of a single dwarf progenitor known as (GSE) that likely fell in early, $\sim 8-10$ Gyr ago \citep{helmi2018, belokurov2018MNRAS_mwdiscandhalo, naidu2020ApJ_stellarhalosub}. 

Unlike the mostly phase-mixed GSE debris of the MW, \katya{M31's halo is $\sim 5$ times more metal-rich than the MW's \citep{gilbert2018}} and contains \katya{more complex} substructure stemming from its more recent merger $\sim 2-3$ Gyr ago \citep[][see also \citealt{Hammer2018}]{dsouzabell2018bm31}. M31's younger merger gives it the kinematic substructure that makes it detectable with our LOESS smoothing variance metric from Sec. \ref{subsec:loess}, making NGC 253 similar to M31 in this regard and reflecting its halo's dynamical youth. 

\subsubsection{Halo AM}
In terms of angular momentum, the overall halo of the MW out to 50 kpc only exhibits weak prograde rotation with $v_\phi = 5-25$ km/s \citep{deason2017MNRAS_spinofMWhalo, brown2010AJ_MWveldisp, king2015ApJMWinnerhaloveldisp, hunt2025NewAR_MWdynamicsreview}, and the most recent data from DESI DR2 reveal that the MW has an  
oblate, disk-aligned, and slightly prograde rotating inner stellar halo ($<20-30$ kpc), though it is primarily dominated by the GSE which may have slight retrograde rotation \citep{helmi2018, belokurov2018MNRAS_mwdiscandhalo, hunt2025NewAR_MWdynamicsreview}. Its radial velocity dispersion is high in the innermost part of the halo $\sim$141 km/s, decreasing to $\sim$100 km/s around r=20 kpc \citep{Bland-Hawthorn2016ARA&A_MWreviewveldsip}. 

M31 has a clear signature of prograde rotation of $\sim150$ km/s at radii of 40-70 kpc along the major axis \citep{ibata2005ApJ_m31accretion} and a relatively high velocity dispersion of $\sim108$ km/s in the inner halo decreasing to $\sim 80$ km/s in the outer halo \citep{gilbert2018}. NGC 253 has similarly high velocity dispersion, though like M31 the presence of substructure makes this hard to measure and generalize. 

Lastly, \cite{beltrand2024_ngc4945} measured the kinematics of NGC 4945's halo at a distance of 34.6 kpc from its center along its major axis in a 1' $\times$ 1' FOV with VLT/MUSE. They find a LOS velocity of $\sim -44$ km/s and a particularly low-velocity dispersion of $42 \pm 22$ km/s. 
\katya{Taking this evidence at face value, NGC 4945's stellar halo appears more like the MW’s than M31’s, with weak counterrotation. While \citet{bell2026OJAp_mergersAM} find that wider-area kinematics at 35 kpc are likely to mirror overall halo AM, we caution that with such a small FOV, it is possible that this measurement reflects substructure instead of the phase-mixed stellar halo. Future wider-area kinematics would be useful in disentangling the two scenarios.}

\katya{Given our separation of NGC 253 stars into \textit{in-situ} and accreted, discussing its halo AM is not straightforward. We detect a halo shell that is offset from the minor-axis along the disk plane of the galaxy that only exhibits a very small rotational signature, which may indicate a weakly or nonrotating halo more akin to the MW or NGC 4945. In parallel, we detect a cluster of highly positive velocity stars along the major axis that we attribute mostly to \textit{in-situ} by comparing with gas velocities, but without concrete evidence from e.g. metallicity, there is a possibility that some of these stars may be part of a phase-mixed halo.} 

\subsection{Caveats and Future Outlook}
We note that one caveat of our work is in our stellar selection. Given the challenge of sky subtraction, it was necessary to select the brightest halo stars we could, which resulted in us choosing only AGB stars to observe rather than a mix of RGB/AGB stars or just RGB as are commonly used for tracing accreted stellar halo populations in nearby galaxies. Though the spatial distributions of AGB and RGB stars are roughly similar, we were more sensitive to tracing out the young/intermediate-age stellar populations present with AGB stars as opposed to generally older RGB populations. Besides different ages, prior to infall, AGB and RGB stars in a progenitor galaxy can show different kinematic and spatial structure (e.g. AGB stars tend to be more centrally concentrated in the disk where there may be active star-formation and with lower velocity dispersion than RGB stars, whose populations can be more kinematically hot and spatially extended to the outer halo). Therefore, we may have seen a lower dispersion or different AM signature or substructures had we been able to get spectra of RGB stars. 

Additionally, we note that though we were able to recover velocities only over a narrow region of $\sim$ 2000\AA, \katya{any future studies could consider obtaining spectra over a wider wavelength region to capture a broader structure of the molecular bands and more prominent absorption lines, though whether this is wholly beneficial is unclear. This may facilitate easier cross correlation or full-spectrum fitting, but would also cause more losses around sky lines if the resulting spectra are lower resolution.}

Though we were only able to recover velocities for 40\% of our total sample, next-generation wide-field multiobject spectrographs on large telescopes would facilitate observations of more stars at fainter magnitudes, letting us build a larger kinematic sample of stars and better understand the nature of NGC 253's halo substructure. Having a larger FOV that covers the entire spatial extent of the inner halo and recovering a larger sample of stellar velocities would enable a more robust measurement of the substructure signal and better understanding of the overall AM and kinematics present. For example, \cite{Greggio2014} find evidence of another overdensity at $\sim 30$ kpc in the north/NW of NGC 253 which they also attribute to a past merger. Spectroscopy of AGB stars associated with this overdensity would provide an important kinematic test and enable a direct comparison between its velocity structure and the SW shell studied in this work. \katya{Additionally, more sensitive spectrographs that are able to reach a higher S/N would let us go one step further and not only derive velocities, but also metallicities and abundances of halo stars. This would provide extra evidence when differentiating between accreted and \textit{in-situ} populations, as well as more robustly identifying stars belonging to different substructures, as these different populations would likely be chemically distinct.} Furthermore, future halo kinematic measurements in other galaxies may be key to understanding the processes behind disk growth and reorientation, especially in halos that show signs of dynamical youth. 

Though there exist alternate tracers of halo kinematics such as integrated light, planetary nebulae \citep{hartke2022A&A_m105PNe,  Bhattacharya2023MNRAS_m31PNe, pulsoni2023A&A_ePNSangularmomentumetg, hartke2025FrASS_PNeaccretion, tsakonas2025A&A_PNeM31}, and GCs \citep{romanowsky2012ApJ_m87GCs, dolfi2021, kruijssen2019MNRAS_MWassemblythruGCs, Fahrion2025A&AM104GCs}, these have mostly been applied to large early type galaxies and have limited sample sizes, population biases, and may not reflect the same overall dynamics as halo stars. For example, \cite{finn2026MNRAS_GCsubstructure} find that kinematic mixing can erase signatures of accretion from GC kinematics alone, making it difficult to distinguish between \textit{in-situ} and \textit{ex-situ} clusters and necessitating the use of a combination of chemo-dynamical properties to link them back to the accretion history of the galaxy. Integrated light kinematics has mostly been done for distant early types that are small enough to observe with small FOV IFUs \citep{Weijmans2009MNRAS.398..561W_IFUstellarhalovelocities, murphy2011ApJ_m87darkmatterhalo,Romanowsky2014ASPC_kinematicsofstreamshalos}, and planetary nebulae may be accreted in different events than GCs \citep[e.g.,][]{coccato2013MNRAS_accretioneventsinETGswPNEandGC} and have differing kinematic signatures. Future systematic studies that are able to survey and compare the complete population of multiple kinematic tracers may paint the most complete picture of the assembly histories of galaxies and enable us to understand how we may cross-validate inferences from each tracer.





\section{Conclusion}\label{sec:conclusion}
This work presents a novel measurement of kinematic substructure in the stellar halo of NGC 253. Using observations of bright AGB stars with medium-resolution spectra from Magellan/M2FS, we perform challenging sky subtraction using \texttt{Skycorr} and PCA and attempt to measure LOS velocities for each star using the cross correlation program \marz. Quantifying how robustly we are able to recover velocities using a set of mock stellar spectra with representative sky-subtraction residuals, we choose a set of recovery criteria and apply these to our observed spectra, recovering individual velocities for 89 stars in our sample. From these individual stars, we calculate a median velocity with respect to NGC 253 of $18_{-16}^{+33}$ km/s with a velocity dispersion of $136_{-16}^{+20}$ km/s, which is in agreement with a measurement of these quantities by \ppxf\ for a single median stack of all these spectra. Both LOESS smoothing and SC identified kinematically distinct parts of NGC 253's stellar halo, with a velocity gradient from negative to positive going from the minor to major-axis, and a SW region with almost-zero velocity. \katya{Comparisons with neutral gas rotation velocities and simulations suggest that the major-axis cluster with distinctly high positive velocities ($v_{med, in-situ} = 167_{-14}^{+16}$ km/s) is likely mostly composed of \textit{in-situ} stars belonging to the extended outer disk of the galaxy, while the modestly negative \urgent{velocity} minor-axis cluster and offset zero-velocity SW cluster likely signify the presence of an accreted population ($v_{med, acc.} = -34_{-12}^{+16}$ km/s). }
This is the first detection of kinematic substructure in a stellar halo outside of the LG, and the fourth-ever measurement of resolved-star halo kinematics in a MW-mass galaxy. The SW zero-velocity region also coincides with a shell structure that was previously identified in integrated light and resolved stars, indicating that NGC 253 likely underwent a recent accretion. Further dynamical modeling may be able to quantify the timing of this merger, but this is outside the scope of this paper.

Because this work was only possible by using the brightest AGB stars due to heavy sky contamination and also benefits from a larger spatial selection to probe the nuances of substructure, we are currently limited to observing the nearest galaxies that also have a significant population of halo AGB stars. As taking spectra of individual stars one-by-one is prohibitive, this type of work also necessitates the use of multiobject spectrographs or IFUs that cover a sufficiently large FOV. In the coming years, 10-30m telescopes such as the ELT, GMT, and TMT as well as wide-field multiobject spectrographs such as Subaru/FMOS, the Wide-Field Spectroscopic Telescope, or VLT/MOONS \citep{gmacs2018SPIE10702E..1XD, fmos2010PASJ...62.1135K, wst2024arXiv240512518B, moons2014SPIE.9147E..0NC} will be game changers for kinematically resolving the stellar halos of nearby galaxies to unprecedented depths and may let us further connect halo velocities to the dynamics and orbit of the most massive progenitor satellites. This work serves as a precursor for future halo kinematic studies in other Local Volume galaxies.


\begin{acknowledgments}
We thank the anonymous referee for the helpful comments, which greatly improved this manuscript. K.G. thanks the University of Michigan Astronomy Department for their gracious funding to support three in-person observing runs to Las Campanas Observatory for this project, 
 as well as all the wonderful support astronomers, telescope operators, and support staff at LCO. She also thanks Sean Johnson for helpful guidance and advice on sky subtraction and spectra stacking, and Rachel Bezanson for a helpful discussion and suggesting \marz\ for velocity fitting. We thank David Sand and Burçin Mutlu-Pakdil for sharing the PISCES catalog with us for optimal sky fiber selection, as well as Marina Rejkuba and Laura Greggio for contributing the VISTA data we used for preparing the observations. We also thank Monica Valluri for helpful discussions. This work was partially supported by the National Science Foundation through grant NSF-AST 2007065 and by NASA under grant No. 80NSSC24K0084 under solicitation NNH22ZDA001NROMAN: Nancy Grace Roman Space Telescope Research and Support Participation Opportunities. A.M. gratefully acknowledges support by FONDECYT Regular grant 1251882 and by the ANID BASAL project FB210003, as well as funding from the HORIZON-MSCA-2021-SE-01 Research and Innovation Programme under the Marie Sklodowska-Curie grant agreement No. 101086388. This research has made use of NASA’s Astrophysics Data System Bibliographic Services. We acknowledge the use of
ChatGPT and Claude to assist with debugging code and for minor revisions to improve the
clarity and readability of the text.
\end{acknowledgments}

\begin{contribution}

K.G. wrote two of the observing proposals used to obtain the data, performed the sky subtraction and ensuing analysis, created all the plots, and wrote the paper. M.M. did the initial image and spectral reduction. E.F.B. initiated the project, wrote the initial observing proposal, did the target selection, ran the TNG-50 analyses, and made significant contributions to the project ideation. E.F.B., R.D., and M.M. contributed to the project ideation and helped with the analysis. K.G., J.B., and M.M. did the observing. All other authors contributed to project execution, writing, review, and editing.

\end{contribution}

%
\facilities{Magellan/M2FS}

\software{\code{Python 3.11} \citep{python} | \code{numpy} \citep{numpy}, \code{matplotlib} \citep{matplotlib}, \code{astropy} \citep{astropy, 2013A&A...558A..33A,2018AJ....156..123A,2022ApJ...935..167A},  \code{scipy} \citep{scipy}, \code{Jupyter} Notebooks \citep{jupyter}, \texttt{pPXF} \citep{ppxf_Cappellari2023};  Skycorr \citep{skycorr2014}, \texttt{Marz} \citep{hinton2016A&C_marz}}


\appendix

\section{Mock Observations} \label{sec:appendix}
This section details the creation and exploration of mock stellar spectra that we used to determine our recovery criteria and velocity measurement error for our stars. For each of the 42 sky-subtracted sky fibers, we make combinations of the following spectral and \marz\ parameters:
\begin{enumerate}
    \item 3 different base \urgent{Archéologie avec Matisse Basée sur les aRchives de l’ESO (AMBRE; \citealt{delaverny2012A&A_ambrespectra}}) template spectra with (T (K), $\log g$)  = [(3500,0.5), (3700,0.5), (3500,1.5)], an [Fe/H] = -1.0, and [$\alpha$/Fe] = 0.4 that were convolved, rebinned, and continuum-adjusted to match the continuum of a median spectrum;
    \item These templates are multiplied by factors of [0.75, 1, 2] to simulate varying the flux of the star. This corresponds to an array of magnitudes for each mock between roughly 23.3 and 22.2 magnitudes in the VISTA Z-band, spanning the range of observed magnitudes of our stars;
    \item Each sky-subtracted sky fiber is the result of coadding with a median or a weighted average as referenced in Sec. \ref{subsec:spectralcoadditionstacking};
    \item The noise of each resulting mock is taken from the noise of the sky-subtracted sky fiber, which set to either be the standard deviation spectrum of that fiber, the standard deviation spectrum with artificially inflated errors as done in Sec. \ref{subsec:spectralcoadditionstacking}, or a uniform standard deviation value of 100 across all wavelengths.
    \item Each spectrum was shifted to a velocity in [500, 200, 0, -200] km/s;
    \item As with the actual data, we limit \marz\ to fit only ranges of $\pm$[1800, 3000, 6000] km/s.
\end{enumerate}

This creates a grid of 27216 spectra with varying sky residuals and S/Ns that we then fit with \marz. We group together spectra that come from the same base mock spectrum that also share the same velocity shift, sky fiber, and magnitudes for a total of 1512 groups (i.e. distinct spectra), and calculate different filters that we can apply to our observed data that can weed out spectra that were not well-fit given different coadding techniques, velocity limits, and noise calculations (i.e. fitting settings, such that each group has 18 different combinations of settings). Since we do not know the actual velocities of each observed star, we turn to metrics based on the spreads in recovered velocity.  We calculate whether:
\begin{enumerate}
    \item the robust standard deviation of the velocities within a group (using the median absolute deviation) was $\leq 50$ km/s
    \item all but two coadd weighted-averaged spectra within a group had recovered velocities that were within $\pm$50 km/s of the median recovered velocity of all coadd weighted-averaged spectra within a group
    \item all but two coadd median spectra within a group had recovered velocities that were within $\pm$50 km/s of the median recovered velocity of all coadd median spectra within a group
    \item all but two spectra in a particular group had recovered velocities that were within $\pm$50 km/s of the median recovered velocity of all spectra in that group.
    \item the absolute value of the difference between the median recovered velocity of the average-weighted spectra and the median spectra was $\leq 100$ km/s
    \item the median recovered velocity of a group was within 400 km/s of the input velocity
\end{enumerate}

\begin{figure}
    \centering
    \includegraphics[width=0.99\linewidth]{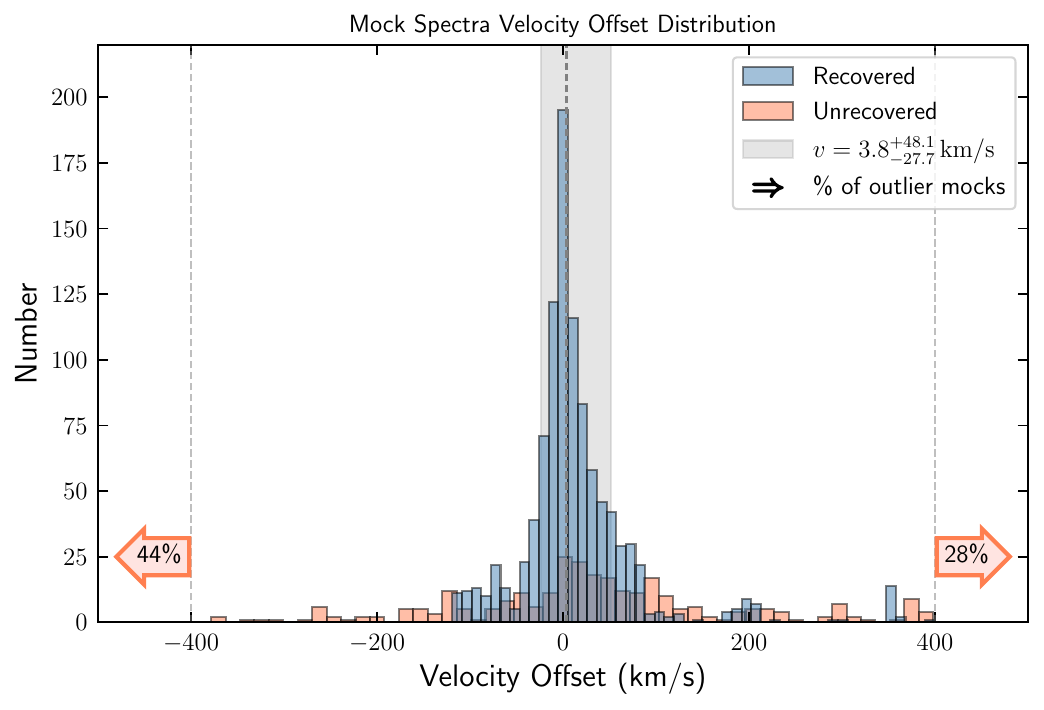}
    \caption{The distribution of velocity offsets (\marz-measured velocity - input velocity) for our mock spectra. The blue histogram shows the velocity offsets of spectra that we deem recovered by our chosen combination of metrics (see text), and the red histogram shows the remaining spectra that were not recovered. Some of our unrecovered spectra had velocity offsets that were much greater than the input velocity -- we visualize only velocity offsets between $\pm400$ km/s and for others, we show arrows showing the percentage of mock spectra within the 
    not recovered (red arrows) distributions that had velocity offsets outside this range. Note that out of the 
    unrecovered spectra, $\sim72\%$ are outliers whose \marz\ -measured velocities were more than 400 km/s from their input velocities.}  
    \label{fig:mock_veloffsets}
\end{figure}

Each logical combination of these two metrics (e.g. 2, 2 AND 3, 2 OR 3, 2 OR 3 AND 5, etc.), determines a set of groups that we deem ``recovered" if they meet all the criteria. To evaluate which set of metrics lets us recover an optimal number of stars,  we calculate the F1 score for each combination. This is given by the harmonic mean of the precision and recall e.g. $F_1 = \frac{2 \times \text{precision} \times \text{recall}}{\text{precision} + \text{recall}}$, where $\text{precision} = \frac{\text{TP}}{{\text{TP} + \text{FP}}}$ and $\text{recall} = \frac{\text{TP}}{{\text{TP} + \text{FN}}}$ and TP = true positives, which are the number of recovered groups with a median velocity within $\pm$50 km/s of the input velocity, FP = false positives, which are the number of recovered groups with median velocity $>50$ km/s of the input velocity (i.e. we deem them recovered even though they should not be), and FN = false negative, which are the number of groups that were \textit{not} recovered despite having median velocity within $\pm$50 km/s of the input velocity (i.e. they should have been recovered). In our case, the F1 score is essentially a measure of the accuracy of \marz\ fitting given different metrics for deeming a star ``recovered" or not, given that we'd like to include as many accurately measured stellar velocities as possible without including too many outliers (i.e. inaccurately measured velocities) or miss too many stars that were accurately measured but not deemed recovered by our metrics. We also look at the total number of groups recovered given each combination of metrics, as we also want to get velocities for as many stars as possible. Based on this analysis, we choose to use the combination of metrics (2 OR 3 OR 5) AND 6, which yielded an $F_1 \sim 0.8$ while recovering $\sim 50\%$ of total groups. Fig \ref{fig:mock_veloffsets} shows the distribution of velocity offsets (\marz-measured velocity - input velocity) for the recovered and unrecovered mocks. 
The median offset for recovered mocks is $3.8_{-27.7}^{+48.1}$ km/s, where the lower/upper error is given by the 16th/84th percentiles of the distribution. 

\bibliography{sample701}{}
\bibliographystyle{aasjournalv7}



\end{document}